\documentclass[a4paper,11pt]{article}
\usepackage[a4paper,left=2.73cm,right=2.7cm,top=3cm,bottom=3.5cm]{geometry}
\usepackage{bbold}
\usepackage{amsfonts,amsmath,amssymb,tabstackengine}
\usepackage{mathtools,slashed}
\usepackage{tensor,mathtools,mathrsfs}
\usepackage[usenames, dvipsnames]{color}
\usepackage{array}
\usepackage{color, colortbl}
\definecolor{Gray}{gray}{0.95}

\usepackage{tikz}
\usetikzlibrary{arrows.meta,calc,decorations.markings,patterns}

\usepackage[colorlinks=true,linktocpage=true,linkcolor=blue,citecolor=blue]{hyperref}
\usepackage{graphicx}
\usepackage{comment}

\addtocontents{toc}{\protect\setcounter{tocdepth}{3}}
\numberwithin{equation}{section}

\usepackage{color, colortbl,xcolor}

\begin{document}

\begin{titlepage}

\thispagestyle{empty}

\begin{center}

{\LARGE \textbf{\mbox{On the branes behind scale-separated AdS$_{3}$ flux vacua}}}

\vspace{40pt}

{\large \bf \'Alvaro Arboleya}$^{(1,2)}$ \,  \large{,} \,  {\large \bf Adolfo Guarino}$^{(1,2)}$

\vspace{8pt}

\,\,\quad\quad {\large \bf Clara Rold\'an-Dom\'inguez}$^{(3)(1,2)}$  \,  \large{,} \, {\large \bf Giuseppe Sudano}$^{(1,2)}$

\vspace{25pt}

{\normalsize  
${}^{1}$ Departamento de F\'isica, Universidad de Oviedo,\\
Avda. Federico Garc\'ia Lorca 18, 33007 Oviedo, Spain.}
\\[3mm]

{\normalsize  
${}^{2}$ Instituto Universitario de Ciencias y Tecnolog\'ias Espaciales de Asturias (ICTEA) \\
Calle de la Independencia 13, 33004 Oviedo, Spain.}
\\[3mm]

{\normalsize
${}^{3}$ Universidade Federal do Esp\'irito Santo\\ 29075-910 Vit\'oria, ES, Brasil.}
\\[10mm]

\texttt{arboleyaalvaro@uniovi.es}  \,\, ,  \,\,
\texttt{adolfo.guarino@uniovi.es}\\
\texttt{clararoldan4@gmail.com} \,\, , \,\, \texttt{gsudano1@gmail.com}  

\vspace{20pt}

\today

\vspace{20pt}

\abstract{\noindent We investigate the brane origin of supersymmetric and scale-separated AdS$_3$ flux vacua arising in a class of type IIB orientifold reductions with G$_2$-structure. First, using the effective three-dimensional supergravity together with explicit uplift formulae, we trace back the $\,F_{(7)}\,$ and $\,F_{(3)}\,$ fluxes responsible for scale separation and identify the type IIB background that is probed by the corresponding D1- and D5-branes. Second, by restoring such D1- and D5-branes, we obtain a type IIB solution that interpolates between the previous background in the asymptotic region and the scale-separated AdS$_3$ flux vacua in the near-horizon region. Third, by working directly in ten dimensions, we construct a codimension-one D1–D5–KK5 intersection whose near-horizon region realises the scale-separated AdS$_3$ flux vacua. Our results provide a higher-dimensional interpretation of these flux vacua in terms of (smeared) brane configurations.}

\end{center}

\end{titlepage}

\tableofcontents

\hrulefill
\vspace{10pt}


\section{Introduction}

Identifying the brane intersection underlying scale-separated anti-de Sitter (AdS) vacua is a crucial step towards establishing their higher-dimensional origin and gaining insight into the structure of the dual conformal field theories (CFT's). By scale-separated AdS vacua we mean type II or 11D supergravity backgrounds of the form $\,\mathrm{AdS}\,\times\,\mathcal{M}\,$ in which the AdS curvature radius is parametrically larger than the characteristic overall size of the internal space $\,\mathcal{M}$.\footnote{A precise assessment of scale separation would require showing the proper decoupling of the Kaluza--Klein (KK) modes, thereby ensuring a lower-dimensional effective description. Since a detailed KK spectral analysis is not available for the general class of AdS$_3$ flux vacua considered in this work, we instead use the decoupling of the characteristic lengths of the internal space as a proxy for scale separation.} To the best of our knowledge, no AdS/CFT pair exhibiting scale separation has yet been established. For instance, in the canonical examples of type IIB on $\,\mathrm{AdS}_5 \times \mathrm{S}^5/{\mathcal{N}=4}\,$ SYM$_4$ \cite{Maldacena:1997re}, 11D supergravity on $\,\mathrm{AdS}_4 \,\times\, \mathrm{S}^7/\mathcal{N}=8\,$ ABJM theory \cite{Aharony:2008ug} (at Chern--Simons (CS) level $k=1$), and massive type IIA on $\,{\mathrm{AdS}_4 \,\times\, \mathrm{S}^6}/{\mathcal{N}=8}\,$ CS-SYM$_3$ \cite{Guarino:2015jca}, the AdS radius and the radius of the internal sphere are of the same order. As a result, none of these AdS/CFT pairs exhibits scale separation. Despite this, supersymmetry ensures the existence of a consistent lower-dimensional supergravity description capturing the low-lying modes of the Kaluza--Klein (KK) spectrum. From a holographic perspective, scale separation is especially interesting because it would correspond to a conformal field theory exhibiting a parametrically large gap between the low-lying spectrum and the tower of operators dual to KK excitations. However, an obstruction to the existence of such a CFT was pointed out in \cite{Bedroya:2025ltj} under the standard assumption that the holographic CFT describes the infrared degrees of freedom of a brane system that decouples from gravity (see \cite{Aharony:2008wz} for a first exploration of the CFT dual to the scale-separated DGKT vacua \cite{DeWolfe:2005uu}).

Still, examples of scale-separated AdS vacua have been proposed within the contexts of type II Calabi--Yau (CY) compactifications and Scherk--Schwarz (SS) reductions in the presence of background fluxes and O-plane/D-brane sources in the smeared limit. These include the massive type IIA AdS$_{4}$ flux vacua of DGKT \cite{DeWolfe:2005uu} (see also \cite{Camara:2005dc}), and the more recent AdS$_{3}$ flux vacua arising in massive type IIA \cite{Farakos:2020phe} (see also \cite{Farakos:2023nms,Farakos:2023wps}) and type IIB \cite{Arboleya:2024vnp} (see also \cite{VanHemelryck:2025qok,Arboleya:2025ocb}) orientifold reductions on manifolds with G$_2$-holonomy \cite{Farakos:2020phe} and (co-closed) G$_2$-structure \cite{Emelin:2021gzx}, respectively.\footnote{Additional examples are the type I scale-separated flux vacua of \cite{Miao:2025rgf}, which are T-dual to the type IIA flux vacua of \cite{Farakos:2020phe,Farakos:2025bwf}, or the heterotic scale-separated flux vacua of \cite{Tringas:2025bwe}, which are S-dual to the type I flux vacua. We refer the reader to \cite{Aikot:2026tzn} for a unified discussion of type IIA, type IIB, type I and heterotic compactifications on seven-dimensional manifolds with G$_2$-structure.} The type IIA scale-separated flux vacua were re-examined in \cite{Apers:2022zjx}  from the perspective of several Swampland conjectures. While the AdS$_4$ DGKT vacua \cite{DeWolfe:2005uu} passed various Swampland-based tests (\textit{e.g.} positivity of the metric over the space of vacua \cite{Palti:2024voy}), a refined version \cite{Buratti:2020kda} of the AdS Distance Conjecture \cite{Lust:2019zwm} was shown to be inconsistent with the massive type IIA AdS$_3$ flux vacua of \cite{Farakos:2020phe}. By contrast, the type IIB scale-separated AdS$_3$ flux vacua constructed in \cite{Arboleya:2024vnp,VanHemelryck:2025qok,Arboleya:2025ocb} have received less attention, possibly because they are rare \cite{Arboleya:2025jko}. Nevertheless, several potential pathologies have already been investigated. For instance, for the non-supersymmetric AdS$_3$ vacua of \cite{Arboleya:2024vnp}, the open-string instabilities anticipated in \cite{Danielsson:2016mtx} on the basis of the AdS Swampland Conjecture \cite{Ooguri:2016pdq} were shown to be absent in \cite{Arboleya:2025lwu}. Another potential issue has recently been raised from a holographic perspective. In \cite{Bobev:2025yxp}, a consistency condition for the existence of a holographic large-$N$ CFT dual to an AdS gravitational effective field theory was proposed, requiring the vanishing of cubic bulk couplings involving scalars dual to operators with extremal (and over-extremal) arrangements of their conformal dimensions, \textit{i.e.} $\,\Delta_i=\Delta_j+\Delta_k\,$. This consistency condition is not satisfied by either the original non-supersymmetric type IIB scale-separated vacua of \cite{Arboleya:2024vnp} or the supersymmetric variants of \cite{VanHemelryck:2025qok}, as recently pointed out in \cite{Revello:2026eqp}, where additional orbifold projections were introduced to eliminate the problematic cubic couplings.\footnote{Such a holographic constraint on cubic couplings holds in DGKT-like scenarios on generic Calabi--Yau three-folds provided $\,h^{2,1}=0\,$ \cite{Revello:2026opc,Revello:2026yla}.} However, we will provide explicit examples of scale-separated type IIB flux vacua where such (over) extremal arrangements do not occur.

This work continues investigating the class of scale-separated type IIB AdS$_3$ flux vacua in \cite{Arboleya:2024vnp,VanHemelryck:2025qok,Arboleya:2025ocb}. These solutions admit a consistent description in terms of three-dimensional $\,\mathcal N=1\,$ gauged supergravity and are supported by a combination of gauge and metric fluxes, \textit{i.e.}, they arise from a SS reduction on a group manifold $\,\mathcal{M}_{7}$. An interesting feature of these vacua is that the fluxes controlling the hierarchy of scales are not constrained by the tadpole cancellation conditions. This observation suggests that these unrestricted fluxes may admit a direct interpretation in terms of would-be colour branes, and raises the broader question of whether the corresponding AdS$_3$ flux vacua can be understood as the near-horizon region of a full brane intersection (including also branes responsible for the restricted fluxes). An interesting strategy for addressing this question is provided by the flux backtracking method introduced in \cite{Apers:2025pon} and further applied to the DGKT AdS$_{4}$ vacua in \cite{Apers:2026lgi}. The basic idea is to exploit the existence of an effective lower-dimensional description together with explicit uplift formulae in order to reconstruct ten-dimensional backgrounds associated with selected sectors of the flux compactification. These ten-dimensional backgrounds can be interpreted as geometries being probed by those D-branes whose charges (or associated fluxes) are unrestricted in the lower-dimensional description. This approach provides a systematic bridge between effective supergravity and higher-dimensional brane setups. An alternative approach is to work directly in ten dimensions and try to construct a full brane intersection whose near-horizon region reproduces the desired AdS flux vacua, as done in \cite{Kounnas:2007dd} for the type IIA AdS$_{4}$ flux vacua of \cite{Derendinger:2004jn,Villadoro:2005cu,DeWolfe:2005uu,Camara:2005dc}.

In this work, we combine the two approaches outlined above to gain new insight into the brane origin of the scale-separated type IIB AdS$_3$ flux vacua of \cite{Arboleya:2024vnp,VanHemelryck:2025qok,Arboleya:2025ocb}. Our analysis proceeds as follows:

\begin{itemize}

\item Firstly we use the effective three-dimensional supergravity description together with the explicit uplift formulae to trace back the unrestricted fluxes responsible for scale separation in the AdS$_3$ vacua. In this way, we reconstruct a ten-dimensional solution that interpolates between a type IIB background (with a strongly-coupled singularity) in the asymptotic region, and the $\,\textrm{AdS}_3 \times \mathcal{M}_{7}\,$ flux vacua of \cite{Arboleya:2024vnp,VanHemelryck:2025qok,Arboleya:2025ocb} in the near-horizon region. This provides the first indication of the brane configuration underlying these flux vacua, with the asymptotic background being naturally interpreted as the geometry probed by the D1- and D5-branes associated with the unrestricted fluxes.

\item Secondly we construct directly in ten dimensions the full (codimension-one) brane intersection consisting of D1-branes, D5-branes, and KK5 monopoles, whose near-horizon region precisely reproduces the scale-separated AdS$_3$ flux vacua of \cite{Arboleya:2024vnp,VanHemelryck:2025qok,Arboleya:2025ocb}. From this perspective, the previous ten-dimensional interpolating solution reconstructed from the lower-dimensional theory arises naturally as a partial near-horizon limit of the full D1-D5-KK5 system.

\end{itemize}

Importantly, the D1- and D5-branes associated with the unrestricted fluxes responsible for scale separation enter the (codimension-one) brane intersection through harmonic functions, and the corresponding source branes at the horizon do not source ten-dimensional Bianchi identities. In contrast, the rest of D5-branes, which are associated with the fluxes entering the tadpole cancellation conditions, are described by non-harmonic functions, and the corresponding brane sources at the horizon contribute to the ten-dimensional Bianchi identities. Non-zero flux-induced tadpoles (particularly those with the sign associated with orientifold planes) were argued in \cite{Tringas:2025uyg} to be essential for achieving scale separation. Nevertheless, they can be restricted to just one type of O-plane/D-brane source in the compactification scheme. In this case, the effective three-dimensional supergravity exhibits an enhancement of supersymmetry from minimal ($\mathcal{N}=1$) to half-maximal ($\mathcal{N}=8$), thereby recovering the $\,\mathcal{N}=8\,$ supergravity models studied in \cite{Arboleya:2024vnp,Arboleya:2025ocb,Arboleya:2025jko}.

The work is organised as follows. In Section~\ref{sec:G2_structure&AdS3_vacua}, we review the class of type IIB orientifold reductions with co-closed G$_2$-structure of relevance for this work. We introduce the effective three-dimensional $\,{\mathcal{N}=1}\,$ supergravity description in terms of a real superpotential, and present the supersymmetric and scale-separated AdS$_3$ flux vacua. In Section~\ref{sec:flux_backtracking}, we apply the flux-backtracking method introduced in \cite{Apers:2025pon,Apers:2026lgi}, and reconstruct the ten-dimensional background featuring a strongly-coupled singularity that is probed by the D1- and D5-branes associated with the unrestricted fluxes responsible for scale separation. Then, by reinstating such D1- and D5-branes, we construct a type IIB solution that interpolates between the previous background and the scale-separated AdS$_{3}$ flux vacua. In Section~\ref{sec:full_intersection}, we present a D1-D5-KK5 intersection and analyse both its asymptotic and near-horizon regions, showing that the latter reproduces the scale-separated AdS$_3$ flux vacua. Section~\ref{sec:discussion} contains a summary of results and future directions to explore. Two appendices complete the work. In Appendix~\ref{app:10D_EOMs} we present the ten-dimensional equations of motion and Bianchi identities of type IIB supergravity in the presence of O-plane and D-brane sources. In Appendix~\ref{app:AdS3_vacua} we collect data of the various families of AdS$_{3}$ flux vacua that are used in the main text to illustrate general results.

\section{AdS$_3$ flux vacua and the flux-brane correspondence}
\label{sec:G2_structure&AdS3_vacua}

In this section we present the type IIB AdS$_{3}$ flux vacua that will be studied throughout this work, and introduce the main ingredients required to analyse their brane realisation. In particular, we discuss the interplay between fluxes, sources, and the effective three-dimensional  supergravity description, thereby setting the stage for the subsequent analysis.

\subsection{The flux compactification picture}

Let us consider type IIB flux compactifications on a very specific class of seven-dimensional group manifolds compatible with a (co-closed) G$_2$-structure, and in presence of background fluxes for the Ramond--Ramond (RR) field strengths $\,F_{(3)}\,$ and $\,F_{(7)}$ (see \textit{e.g.} \cite{Emelin:2021gzx,Arboleya:2024vnp,VanHemelryck:2025qok,Aikot:2026tzn}).

\subsubsection*{10D geometry and dilaton}

We start with an (unwarped) product Ansatz for the type IIB ten-dimensional metric (in string frame) of the form
\begin{equation}
\label{10D_metric}
ds_{10}^2 = \tau^{-2} \, ds_{3}^2 + \rho^2 \, d\tilde{s}_7^2 \ ,
\end{equation}
with $\,ds_{3}^2=g_{\mu\nu} \, dx^{\mu} \, dx^{\nu}\,$ being the line elements of the $D=3$ external spacetime, and with
\begin{equation}
\label{ds7_metric}
d\tilde{s}_7^2 = \sum_{m=1}^{6} \ell_{m}^2 \, \left(\eta^m\right)^{2} + \ell_{7}^2 \, \left(\eta^7\right)^{2} \ ,
\end{equation}
being the one of a seven-dimensional internal space (group manifold) with unit volume, \textit{i.e.} $\,\ell_{7}^{2}=(\ell_{1} \cdots \ell_{6})^{-2}$. The above Ansatz comprises eight real and $x$-dependent scalar fluctuations, commonly referred to as \textit{moduli}, which we will collectively denote
\begin{equation}
\label{dilatons}
\phi^{A}=\{\tau,\, \rho,\, \ell_1,\,\ell_2,\,\ell_3,\,\ell_4,\,\ell_5\,,\ell_6\} \ .
\end{equation}
The two scalars $\,\tau\,$ and $\,\rho\,$ in (\ref{10D_metric}) are universal moduli and determine the string coupling $\,g_{s}\,$ and the internal volume of the internal space as
\begin{equation}
\label{gs&vol7}
g_{s}^{2} = e^{2\Phi} =  \frac{\textrm{vol}_7}{\tau} 
\hspace{6mm} , \hspace{6mm} 
\textrm{vol}_7 = \rho^7 \ .
\end{equation}

\subsubsection*{A simple seven-dimensional group manifold}

We will focus on a particular seven-dimensional group manifold, $\,\mathcal{M}_{7}$, previously considered in \cite{Arboleya:2024vnp,VanHemelryck:2025qok}, whose isometry generators satisfy the algebra
\begin{equation}
\label{algebra_internal_space}
\begin{array}{c}
\left[ X_{2}, X_{7} \right]  = \omega_{1} \, X_{1}
\hspace{6mm} , \hspace{6mm}
\left[ X_{1},X_{7} \right] = - \omega_{2} \, X_{2}  \ , \\[2mm]
\left[ X_{4}, X_{7} \right]  = \omega_{3} \, X_{3}
\hspace{6mm} , \hspace{6mm}
\left[ X_{3},X_{7} \right] = - \omega_{4} \, X_{4}  \ , \\[2mm]
\left[ X_{6}, X_{7} \right]  = \omega_{5} \, X_{5}
\hspace{6mm} , \hspace{6mm}
\left[ X_{5},X_{7} \right] = - \omega_{6} \, X_{6} \ ,
\end{array}
\end{equation}
with constant and positive metric fluxes $\,\omega_{a} > 0\,$ ($a=1,3,5$) and $\,\omega_{i} > 0\,$ ($i=2,4,6$).\footnote{The particular choice $\,\omega_{1} = \omega_{3} = \omega_{5} \equiv \omega_{1}\,$ and  $\,\omega_{2} = \omega_{4} = \omega_{6}\equiv \omega_{2}\,$ was originally found in \cite{Arboleya:2024vnp} to be compatible with the existence of scale-separated AdS$_{3}$ flux vacua. An alternative choice $\,\omega_{2} = \omega_{1}\,$, $\,\omega_{4} = \omega_{3}\,$ and $\,\omega_{6} = \omega_{5}\,$ was subsequently considered in \cite{VanHemelryck:2025qok}. Although reaching similar conclusions regarding the scale separation of the overall internal volume, that analysis also highlighted potential concerns related to the decoupling of the KK modes. Be that as it may, the three-dimensional supergravity to be introduced in Section~\ref{sec:3D_supergravity_model} remains being a consistent truncation of type IIB supergravity.} Equivalently, the seven Maurer--Cartan basis elements $\,\{\eta^{a},\eta^{i},\eta^{7}\}\,$ on the group manifold (\ref{ds7_metric}) obey the structure equations
\begin{equation}
\label{Maurer-Cartan_eq}
d\eta^{a} + \omega_{a} \, \eta^{i} \wedge \eta^{7} = 0
\hspace{10mm} \textrm{ and } \hspace{10mm}
d\eta^{i} - \omega_{i} \, \eta^{a} \wedge \eta^{7} = 0 \ ,
\end{equation}
with pairs $\,(a,i)=\{(1,2),(3,4),(5,6)\}$. The system of equations in (\ref{Maurer-Cartan_eq}) can be integrated to obtain the one-form basis elements
\begin{equation}
\label{one-form_integrated_general}
\begin{array}{rcl}
\eta^{a} &=& \phantom{-} \cos\left[ \sqrt{\omega_{a} \, \omega_{i}} \, x^{7}\right] \, dx^{a} + \sqrt{\dfrac{\omega_{a}}{\omega_{i}}} \sin\left[  \sqrt{\omega_{a} \, \omega_{i}} \, x^{7}\right] \,  dx^{i} \ , \\[4mm]
\eta^{i} &=& - \sqrt{\dfrac{\omega_{i}}{\omega_{a}}} \sin\left[  \sqrt{\omega_{a} \, \omega_{i}} \, x^{7}\right] \, dx^{a} + \cos\left[  \sqrt{\omega_{a} \, \omega_{i}}\, x^{7}\right] \, dx^{i} \ , \\[4mm]
\eta^{7} &=& dx^{7} \ .
\end{array}
\end{equation}
The brackets in (\ref{algebra_internal_space}) describe a $2$-step solvable algebra, and the corresponding solvmanifold can be viewed as three copies of the three-dimensional solvmanifold E$_2$ sharing a common generator $\,X_{7}$. In the limiting cases $\,\omega_{a} \rightarrow 0\,$ or $\,\omega_{i} \rightarrow 0$, the group manifold degenerates into a seven-dimensional $2$-step nilmanifold labeled 37A in \cite{Gong1998}. This nilmanifold will play a central role in the AdS$_3$ flux vacuum examples discussed in Sections \ref{sec:example_KK1} and \ref{sec:example_KK2}.

\subsubsection*{Co-closed G$_{2}$-structure and background fluxes}

The internal seven-dimensional solvmanifold in (\ref{ds7_metric}) with (\ref{Maurer-Cartan_eq})-(\ref{one-form_integrated_general}) is compatible with a co-closed G$_2$-structure. This is specified by the G$_2$-invariant three-form and four-form
\begin{equation}
\begin{array}{rcll}
\label{G2-inv_forms}
\varphi_{(3)} &=& \eta^{127}  +  \eta^{347} + \eta^{567}  - \eta^{136} - \eta^{235} -  \eta^{145}   +  \eta^{246} & , \\[2mm]
\varphi_{(4)} &=& \eta^{3456}  +  \eta^{1256} + \eta^{1234} -  \eta^{2457}  -  \eta^{1467} -  \eta^{2367}  +  \eta^{1357} & ,
\end{array}
\end{equation}
such that $\,d\varphi_{(4)}=0$, where we have introduced the shorthand notation $\,\eta^{127} \equiv \eta^{1} \wedge \eta^{2} \wedge \eta^{7}$, etc. The G$_{2}$-invariant three- and four-forms in (\ref{G2-inv_forms}) can be used to turn on background gauge fluxes for the RR field strengths $\,F_{(3)}\,$ and $\,F_{(7)}$. These are given by
\begin{equation}
\begin{array}{rcl}
\label{F3,F7_fluxes}
F_{(3)} &=& f_{31} \, \eta^{127}  +  f_{32} \, \eta^{347} + f_{33} \, \eta^{567}  - f_{34} \, \eta^{136} - f_{35} \, \eta^{235} -  f_{36} \, \eta^{145}   +  f_{37} \, \eta^{246} \ , \\[2mm]
F_{(7)} &=& f_{7} \, \eta^{1234567} \ ,
\end{array}
\end{equation}
with a set of \textit{constant} flux parameters $\,(f_{31},\ldots,f_{37})\,$ and $\,f_{7}$. This completes the setup required for the explicit dimensional reduction, to which we now turn.

\subsubsection{The $\mathcal{N}=1$, $D=3$ effective supergravity}
\label{sec:3D_supergravity_model}

Details of the dimensional reduction of type IIB supergravity on seven-dimensional group manifolds can be found in many references (see \textit{e.g.} \cite{Emelin:2021gzx} for co-closed G$_{2}$-orientifolds). Throughout this work, we follow the notation and conventions in Appendix~A of \cite{Arboleya:2025jko}. Performing the dimensional reduction explicitly, including the Dirac--Born--Infeld (DBI) terms for the brane sources in Table~\ref{tab:sources}, gives rise to an $\,\mathcal{N}=1\,$, $\,D=3\,$ supergravity model with an action
\begin{equation}
\label{S_3D}
S_{\text{3D}} \,\,=\,\, \frac{1}{2\kappa^2}\int dx^3\sqrt{-g}\,\Big(R+ L_{\textrm{kin}} - V\,\Big) \ .
\end{equation}
The kinetic terms for the scalar fields in (\ref{dilatons}) turn out to be of the form
\begin{equation}
\label{Kinetic_matrix}
L_{\textrm{kin}} = - K_{AB}(\partial\phi^A)(\partial\phi^B)  = - \frac{(\partial\tau)^2}{\tau^2} - 7 \frac{(\partial\rho)^2}{\rho^2} - \sum_{m,n=1}^{6}(1+\delta_{mn})\frac{(\partial \ell_m)(\partial \ell_n)}{\ell_m\,\ell_n}  \ ,
\end{equation}
whereas the scalar potential $\,V\,$ can be obtained from a real superpotential $\,W\,$ using the $\,\mathcal{N}=1\,$ formula
\begin{equation}
\label{V_from_W}
V = K^{AB} \, \left(\frac{\partial W}{\partial\phi^A}\right) \, \left(\frac{\partial W}{\partial\phi^B}\right) - 2 \, W^2 \ ,
\end{equation}
where $\,K^{AB}\,$ is the inverse of the kinetic matrix in (\ref{Kinetic_matrix}). For the metric fluxes in (\ref{Maurer-Cartan_eq}) and the gauge fluxes in (\ref{F3,F7_fluxes}) one finds the superpotential
\begin{equation}
\label{W_model}
\begin{array}{rcl}
2 \, g^{-1} \, W &=& \dfrac{f_{7}}{\tau^{\frac{3}{2}} \, \rho^{\frac{7}{2}}}  + \dfrac{\rho^{\frac{1}{2}}}{\tau^{\frac{3}{2}}} \Big( 
  f_{31} \, \ell_{3456} 
+ f_{32} \, \ell_{1256} 
+ f_{33} \, \ell_{1234} 
+ \dfrac{f_{34}}{\ell_{136}}
+ \dfrac{f_{35}}{\ell_{352}}
+ \dfrac{f_{36}}{\ell_{514}}
+ \dfrac{f_{37}}{\ell_{246}}
\Big) \\[4mm]
& + & 
\dfrac{1}{\tau \, \rho} 
\Big[
 \big(\omega_{1} \, \ell_{1}^{2} +  \omega_{2} \, \ell_{2}^{2} \big) \, \ell_{3456}
+
\big(\omega_{3} \, \ell_{3}^{2} + \omega_{4} \, \ell_{4}^{2} \big) \, \ell_{1256}

+\big(\omega_{5} \, \ell_{5}^{2} + \omega_{6} \, \ell_{6}^{2} \big) \, \ell_{1234}
\Big]  \ ,
\end{array}
\end{equation}
where we have introduced the shorthand notation $\,\ell_{3456} \equiv \ell_{3}\ell_{4}\ell_{5}\ell_{6}$, $\,\ell_{136} \equiv \ell_{1}\ell_{3}\ell_{6}$, etc. Then, plugging (\ref{W_model}) into (\ref{V_from_W}) produces the scalar potential
\begin{equation}
\label{V_N1}
\begin{array}{rcl}
2 \, g^{-2} \, V &=& \dfrac{f_{7}^{2}}{\tau^{3} \, \rho^{7}} \\[4mm]
& + & \dfrac{\rho}{\tau^{3}} \left[ \,\,
  f_{31}^{2} \, \ell_{3456}^{2} 
+ f_{32}^{2} \, \ell_{1256}^{2} 
+ f_{33}^{2} \, \ell_{1234}^{2} 
+ \dfrac{f_{34}^{2}}{\ell_{136}^{2}}
+ \dfrac{f_{35}^{2}}{\ell_{352}^{2}}
+ \dfrac{f_{36}^{2}}{\ell_{514}^{2}}
+ \dfrac{f_{37}^{2}}{\ell_{246}^{2}} \,\, 
\right] \\[4mm]
& + & 
\dfrac{1}{\tau^{2} \, \rho^{2}} \,\,
\Big[
\,\, \big(\omega_{1} \, \ell_{1}^{2} - \omega_{2} \, \ell_{2}^{2} \big)^{2} \, \ell_{3456}^{2} 
+
\big(\omega_{3} \, \ell_{3}^{2} -  \omega_{4} \, \ell_{4}^{2} \big)^{2} \, \ell_{1256}^{2} 
+
\big(\omega_{5} \, \ell_{5}^{2} -  \omega_{6} \, \ell_{6}^{2} \big)^{2} \, \ell_{1234}^{2} 
\,\, \Big] \\[4mm]
& - &
\dfrac{2}{\tau^{\frac{5}{2}} \, \rho^{\frac{1}{2}}}  \,\, \Big[ \,\, \big( \omega_{3} \, f_{35} + \omega_{1} \, f_{36} + \omega_{6} \, f_{37} \big) \, \ell_{136} +  \big( \omega_{3} \, f_{34} + \omega_{5} \, f_{36} +\omega_{2} \, f_{37} \big) \, \ell_{235} \\[4mm]
& &
\quad\quad\quad 
+ \, \big( \omega_{1} \, f_{34} + \omega_{5} \, f_{35} + \omega_{4} \, f_{37} \big) \, \ell_{145} 
+ \big( \omega_{2} \, f_{35} + \omega_{4} \, f_{36} + 
\omega_{6} \, f_{34} \big) \, \ell_{246} \,\,  \Big] \ .
\end{array}
\end{equation}
The first and second lines of (\ref{V_N1}) account for the dimensional reduction of the $|F_{(7)}|^{2}$ and $|F_{(3)}|^{2}$ terms in the ten-dimensional bulk action, respectively. The third line corresponds to the scalar curvature of the seven-dimensional internal solvmanifold in (\ref{Maurer-Cartan_eq}). The last two lines stem from the dimensional reduction of the DBI action for the $\widetilde{\textrm{O}5}$/$\widetilde{\textrm{D}5}^{(4,5,6,7)}$ sources in Table~\ref{tab:sources}. We will elaborate more on these brane sources in Section~\ref{sec:source_branes}.

In the following, we set the three-dimensional gauge coupling to $\,g=1\,$ without loss of generality. The dependence on $\,g\,$ can be restored at any stage by rescaling all flux parameters according to $\,f \rightarrow g \, f\,$ and $\,\omega \rightarrow g \, \omega\,$ (see \textit{e.g.} (\ref{W_model}) and (\ref{V_N1})).

\subsubsection{Supersymmetric AdS$_{3}$ flux vacua}

The $\,\mathcal{N}=1\,$ superpotential in (\ref{W_model}) possesses extrema which correspond to supersymmetric AdS$_{3}$ vacuum solutions of the three-dimensional supergravity. Expressing for simplicity the fluxes in terms of the moduli fields, the most general solution to the extremisation conditions $\,\partial_{\phi^{A}}W=0\,$ takes the form
\begin{equation}
\label{general_AdS3_1}
f_{7} =  - \dfrac{\tau^{\frac{1}{2}} \, \rho^{\frac{5}{2}}}{3 \, \ell_{7}} \, \Omega 
\hspace{4mm} , \hspace{3mm}
f_{31} = - \dfrac{\tau^{\frac{1}{2}}}{3 \, \rho^{\frac{3}{2}}} \, \ell_{1} \, \ell_{2} \, \Omega
\hspace{3mm} , \hspace{3mm}
f_{32}  =  - \dfrac{\tau^{\frac{1}{2}}}{3 \, \rho^{\frac{3}{2}}} \, \ell_{3} \, \ell_{4} \, \Omega
\hspace{3mm} , \hspace{3mm}
f_{33} =  - \dfrac{\tau^{\frac{1}{2}}}{3 \, \rho^{\frac{3}{2}}}
\, \ell_{5} \, \ell_{6} \,  \Omega \ , 
\end{equation}
where $\,\ell_{7} \equiv (\ell_{1} \, \ell_{2} \, \ell_{3} \, \ell_{4} \, \ell_{5} \, \ell_{6})^{-1}$, together with
\begin{equation}
\label{general_AdS3_2}
\begin{array}{lcr}
f_{34}  &=&  \dfrac{\tau^{\frac{1}{2}}}{3 \, \rho^{\frac{3}{2}}}
\, \dfrac{\ell_{1} \, \ell_{3} \, \ell_{6}}{\ell_{7}}
\Big[ 2 \left( \Omega_{1} + \Omega_{3}  + \Omega_{6} \right)  - \left( \Omega_{2} + \Omega_{4} + \Omega_{5} \right)  \Big] \ , \\[2mm]
f_{35} & = &  \dfrac{\tau^{\frac{1}{2}}}{3 \, \rho^{\frac{3}{2}}}
\, \dfrac{\ell_{2} \, \ell_{3} \, \ell_{5}}{\ell_{7}}
\Big[ 2 \left( \Omega_{2}  + \Omega_{3} + \Omega_{5} \right) - \left( \Omega_{1} + \Omega_{4} + \Omega_{6}  \right)  \Big] \ , \\[2mm]
f_{36} & = &  \dfrac{\tau^{\frac{1}{2}}}{3 \, \rho^{\frac{3}{2}}}
\, \dfrac{\ell_{1} \, \ell_{4} \, \ell_{5}}{\ell_{7}}
\Big[ 2 \left( \Omega_{1} + \Omega_{4} + \Omega_{5} \right) 
- \left( \Omega_{2} + \Omega_{3} + \Omega_{6} \right)  \Big] \ , \\[2mm]
f_{37} & = &  \dfrac{\tau^{\frac{1}{2}}}{3 \, \rho^{\frac{3}{2}}}
\, \dfrac{\ell_{2} \, \ell_{4} \, \ell_{6}}{\ell_{7}}
\Big[ 2 \left( \Omega_{2} + \Omega_{4} + \Omega_{6} \right) 
- \left( \Omega_{1} + \Omega_{3} + \Omega_{5} \right)  \Big] \ .
\end{array}
\end{equation}
In order to shorten the expressions in (\ref{general_AdS3_1})-(\ref{general_AdS3_2}) we have introduced the positive definite -- recall that $\,\omega_{a},\omega_{i} >0\,$ in (\ref{Maurer-Cartan_eq}) -- scalar-dressed metric fluxes
\begin{equation}
\Omega_{1} \equiv \omega_{1} \dfrac{\ell_{1}}{\ell_{2}}
\,\,\,\, , \,\,\,\,
\Omega_{2} \equiv \omega_{2} \dfrac{\ell_{2}}{\ell_{1}}
\,\,\,\, , \,\,\,\,
\Omega_{3} \equiv \omega_{3} \dfrac{\ell_{3}}{\ell_{4}}
\,\,\,\, , \,\,\,\,
\Omega_{4} \equiv \omega_{4} \dfrac{\ell_{4}}{\ell_{3}}
\,\,\,\, , \,\,\,\,
\Omega_{5} \equiv \omega_{5} \dfrac{\ell_{5}}{\ell_{6}}
\,\,\,\, , \,\,\,\,
\Omega_{6} \equiv \omega_{6} \dfrac{\ell_{6}}{\ell_{5}}
\end{equation}
and introduced the total sum 
\begin{equation}
\Omega \equiv \sum_{m=1}^{6} \Omega_{m} > 0 \ .
\end{equation}
Note that only $\,\Omega\,$ appears in (\ref{general_AdS3_1}) whereas four different linear combinations of $\,\Omega_{m}\,$ enter (\ref{general_AdS3_2}). A quick inspection of (\ref{general_AdS3_1}) and (\ref{general_AdS3_2}) then shows that their structure is governed by the structure of the G$_{2}$-invariant forms in (\ref{G2-inv_forms}). The above relations (\ref{general_AdS3_1})-(\ref{general_AdS3_2}) can be understood as follows. Given a choice of metric fluxes $(\omega_{a},\omega_{i})$ and a supersymmetric AdS$_3$ vacuum specified by the vacuum expectation values (VEVs) of the moduli fields, the corresponding gauge flux configuration can be obtained by evaluating the right-hand sides of \eqref{general_AdS3_1}-\eqref{general_AdS3_2}. This uniquely determines the gauge fluxes (which must be quantised) supporting the AdS$_3$ vacuum. A non-exhaustive exploration of the AdS$_3$ flux vacua (\ref{general_AdS3_1})-(\ref{general_AdS3_2}) indicates that the normalised mass spectrum of scalar fluctuations generally gives rise to irrational conformal dimensions ($\Delta$'s) for the corresponding operators in the putative dual CFT$_2$. However, for sufficiently symmetric flux configurations, such as those to be considered in this work, the conformal dimensions can become integer-valued. This feature was already noted in \cite{Arboleya:2024vnp,Arboleya:2025ocb}, and closely parallels analogous observations originally made for DGKT vacua \cite{Conlon:2021cjk,Apers:2022tfm}.

Specific examples belonging to the general class of supersymmetric AdS$_{3}$ flux vacua in (\ref{general_AdS3_1})-(\ref{general_AdS3_2}) have already appeared in the literature \cite{VanHemelryck:2025qok,Arboleya:2025ocb}. Importantly, after making a choice for the fluxes $(f_{34}\,,\,f_{35}\,,\,f_{36}\,,\,f_{37})$ in (\ref{general_AdS3_2}), the moduli in (\ref{dilatons}) can be stabilised with arbitrary flux parameters
\begin{equation}
\label{fluxes_negative}
f_{7} \,<\, 0
\hspace{10mm} \textrm{ and } \hspace{10mm}
\left(f_{31}  \,,\, f_{32}  \,,\,  f_{33} \right) \,<\, 0 \ ,
\end{equation}
and the vacuum energy takes the form
\begin{equation}
\label{vacuum_energy_general}
|V_{0}| \sim  \dfrac{1}{\left(f_{7} \, f_{31} \, f_{32} \, f_{33}\right)^{2}}  \ .
\end{equation}
The value $\,V_{0}<0\,$ at a given AdS$_{3}$ flux vacuum determines its radius as $\,L^2 \equiv \frac{2}{|V_{0}|}$, and the central charge of the would-be dual CFT$_{2}$ is \cite{VanHemelryck:2025qok}
\begin{equation}
\label{central_charge}
c \sim L \sim f_{7} \, f_{31} \, f_{32} \, f_{33}  \ .
\end{equation}
Note that $\,c\,$ is linear in the product of the fluxes (or D$1$ and D$5^{(1,2,3)}$ brane numbers) as in the standard D1-D5 system \cite{Brown:1986nw,Strominger:1996sh}.

\subsubsection*{Weak-coupling regime and hierarchy of scales}

The supersymmetric AdS$_{3}$ flux vacua in (\ref{general_AdS3_1})-(\ref{general_AdS3_2}) are compatible with a parametrically-controlled regime of weak coupling, large internal volume, and scale separation between the external spacetime and the (overall size of) internal space by appropriately scaling the unrestricted flux parameters $\,(f_{31},f_{32},f_{33})\,$ and $\,f_{7}$. The characteristic size of the internal space ($L_{7}$) and the string coupling constant are given by
\begin{equation}
L_{7} \equiv (\textrm{vol}_{7})^{\frac{1}{7}} \sim \left( \dfrac{f_{7}^{7}}{f_{31} \, f_{32} \, f_{33}} \right)^\frac{1}{28}
\hspace{6mm} \textrm{ and } \hspace{6mm}
g_{s}^{2} \sim \left( \dfrac{f_{7}}{f_{31} \, f_{32} \, f_{33}} \right) \ ,
\end{equation}
whereas the characteristic length of the three-dimensional external spacetime ($L_{3}$) reads
\begin{equation}
L_{3} \equiv \tau^{-1} \, L \sim \left(f_{7} \, f_{31} \, f_{32} \, f_{33} \right)^\frac{1}{4} \ . 
\end{equation}
Taking, as a simple example, integer-valued gauge fluxes of the form
\begin{equation}
\label{scalings_N}
|f_{31}|\sim N^{a_{1}}\,\,\,,\,\,\, |f_{32}|\sim N^{a_{2}} \,\,\,,\,\,\, |f_{33}| \sim N^{a_{3}}
\hspace{10mm} \textrm{ and } \hspace{10mm}
|f_{7}| \sim N^{b} \ ,
\end{equation}
with $\,N \in \mathbb{N}^{+}\,$ being a scaling parameter to be taken in the large-$N$ limit, and with positive definite exponents $\, a_{1},a_{2},a_{3},b > 0$, one finds that
\begin{equation}
\label{scaling_general}
g_{s}^{2} \sim N^{b-a} 
\hspace{10mm} , \hspace{10mm}
L_{7} \sim N^{\frac{7b-a}{28}}
\hspace{10mm} , \hspace{10mm}
\dfrac{L_{7}}{L_{3}} \sim N^{-\frac{2a}{7}} \ ,
\end{equation}
with $\,a \equiv a_{1}+a_{2}+a_{3}$. Therefore, the desired regime is achieved in the limit $\,N \gg 1 \,$ if 
\begin{equation}
\label{decoupling_a_b}
\frac{a}{7} < b < a \ . 
\end{equation}

\subsubsection*{Decoupling of KK modes?}

A precise assessment of scale separation requires demonstrating that the KK modes decouple appropriately, thereby ensuring a lower-dimensional effective description. In the case of the solvmanifold in (\ref{Maurer-Cartan_eq}), it was argued in \cite{VanHemelryck:2025qok} that the characteristic size of the $7^{\textrm{th}}$ radius does not decouple from the external spacetime, \textit{i.e.} $\, \rho\,\ell_{7} \sim L_{3}\,$ in the geometry (\ref{10D_metric}), thus putting at risk the validity of the SS reduction as an effective field theory.

However the situation improves when setting $\,\omega_{i}=0$, in which case the solvmanifold specified by (\ref{Maurer-Cartan_eq}) degenerates into a nilmanifold. Setting $\,\omega_{i}=0\,$  removes the flux parameter $\,f_{37}\,$ from the tadpole cancellation conditions (see eq.(\ref{Tadpole_O5/D5_4567}) below) so that $\,f_{37}\,$ becomes unrestricted and can be rescaled independently of the $\,(f_{31},f_{32},f_{33})\,$ and $\,f_{7}\,$ fluxes in (\ref{scalings_N}). Taking as before integer-valued gauge fluxes of the form
\begin{equation}
\label{scalings_N_KK1}
|f_{31}|\sim N^{a_{1}}\,\,\,,\,\,\, |f_{32}|\sim N^{a_{2}} \,\,\,,\,\,\, |f_{33}| \sim N^{a_{3}} \,\,\,\,\,\,\,,\,\,\,\,\,\,\, |f_{37}| \sim N^{c}
\hspace{8mm} \textrm{ and } \hspace{8mm}
|f_{7}| \sim N^{b} \ ,
\end{equation}
with positive definite exponents, one now finds (see Appendix~\ref{sec:AdS3_KK_1} for more details)
\begin{equation}
g_{s}^{2} \sim N^{b-(a+c)} 
\hspace{10mm} , \hspace{10mm}
L_{7} \sim N^{\frac{7b-(a+c)}{28}}
\hspace{10mm} , \hspace{10mm}
\dfrac{L_{7}}{L_{3}} \sim N^{-\frac{2(a+c)}{7}} \ ,
\end{equation}
which generalises the result in (\ref{scaling_general}).
A weakly-coupled regime with large internal volume and scale separation between this and the external spacetime is achieved in the limit $\,N \gg 1 \,$ if 
\begin{equation}
\frac{a+c}{7} < b < a+c \ .
\end{equation}
An evaluation of the seven radii in the internal geometry (\ref{10D_metric})-(\ref{ds7_metric}) gives
\begin{equation}
\begin{array}{llll}
\label{rescalings_1-6_KK1}
\rho \, \ell_{1} \sim N^{\frac{a_{1}}{2}+\frac{b-(a+c)}{4}}
&\hspace{10mm} , \hspace{10mm} &
\dfrac{\rho \, \ell_{1}}{L_{3}} \sim N^{\frac{a_{1}-(a+c)}{2}} & , \\[4mm]
\rho \, \ell_{2} \sim N^{\frac{a_{1}}{2}+\frac{b-(a-c)}{4}}
& \hspace{10mm} , \hspace{10mm} & 
\dfrac{\rho \, \ell_{2}}{L_{3}} \sim N^{\frac{a_{1}-a}{2}} & ,
\end{array}
\end{equation}
and similarly for $\,(\rho \, \ell_{3} \, , \, \rho \, \ell_{4})\,$ if replacing $\,a_{1} \leftrightarrow a_{2}\,$; and also for $\,(\rho \, \ell_{5} \,,\, \rho \, \ell_{6})\,$ if replacing $\,a_{1} \leftrightarrow a_{3}\,$. In addition, there is also
\begin{equation}
\label{rescalings_7_KK1}
\rho \, \ell_{7} \sim N^{\frac{b+(a-c)}{4}}
\hspace{10mm} , \hspace{10mm} 
\dfrac{\rho \, \ell_{7}}{L_{3}} \sim N^{-\frac{c}{2}} \ . 
\end{equation}
Setting $\,c=0\,$ in (\ref{rescalings_7_KK1}) yields $\,L_{3} \sim \rho \, \ell_{7}\,$ as originally observed in \cite{VanHemelryck:2025qok} for the case of the solvmanifold which, indeed, does not allow for a rescaling of $\,f_{37}\,$ as in (\ref{scalings_N_KK1}). However, for the nilmanifold, we can take $\,c>0\,$ so that $\,L_{3} \gg \rho \, \ell_{7}\,$ when $\,N \gg 1\,$. A working instance that yields $\,g_{s}^{2} \ll 1$, $\,\rho \, \ell_{m} \gg 1\,$ and $\,L_{3} \gg \rho \, \ell_{m}\,$ (for all $\,m=1,\ldots,7$) is given by $\,a_{1}=a_{2}=a_{3}=\frac{b}{3}=c=1$. The resulting scalings in (\ref{rescalings_1-6_KK1})-(\ref{rescalings_7_KK1}) are anisotropic, in agreement with the proposed obstruction to achieving parametric scale separation in isotropic compactifications of \cite{Tringas:2025uyg}. All in all, and despite a proper KK spectroscopy analysis for the nilmanifold remains to be done, we are optimistic that the KK modes will decouple, giving rise to a valid effective field theory.

\subsection{The brane intersection picture}

According to the general flux/brane correspondence of \cite{Kounnas:2007dd}, an AdS flux vacuum may be interpreted as the near-horizon region of a codimension-one domain-wall solution describing intersecting brane configurations in the presence of sources. In this correspondence, the domain-wall branes are replaced, in the near-horizon region, by the corresponding constant fluxes supporting the AdS vacuum, while the source branes manifest themselves through the tadpole cancellation conditions satisfied by the flux background.

\begin{table}[t]
\centering
\renewcommand{\arraystretch}{1.5}
\begin{tabular}{|c||cc||c||cc|cc|cc||c|}
\hline
 & $dt$ & $dx$ & $dy$ & $\eta^{1}$  & $\eta^{2}$ & $\eta^{3}$  & $\eta^{4}$ & $\eta^{5}$ & $\eta^{6}$ & $\eta^{7}$  \\
\hline
\hline
$\widetilde{\textrm{O}5}$/$\widetilde{\textrm{D}5}^{(1)}$   & $\times$ & $\times$ & $\times$  & $\times$  & $\times$  & &  &  &  & $\times$  \\ 
$\widetilde{\textrm{O}5}$/$\widetilde{\textrm{D}5}^{(2)}$    & $\times$ & $\times$ & $\times$  &  &  & $\times$  & $\times$   &  &  & $\times$  \\
$\widetilde{\textrm{O}5}$/$\widetilde{\textrm{D}5}^{(3)}$   & $\times$ & $\times$ &  $\times$ &   &   & &  & $\times$  & $\times$  & $\times$  \\ 
\hline
\hline
$\widetilde{\textrm{O}5}$/$\widetilde{\textrm{D}5}^{(4)}$ & $\times$ & $\times$ & $\times$ & $\times$ &  & $\times$ &  &  & $\times$  &  \\
$\widetilde{\textrm{O}5}$/$\widetilde{\textrm{D}5}^{(5)}$ & $\times$ & $\times$ & $\times$ &  & $\times$  & $\times$ &  & $\times$ &  &  \\ 
$\widetilde{\textrm{O}5}$/$\widetilde{\textrm{D}5}^{(6)}$ & $\times$ & $\times$ & $\times$ & $\times$ &  &  & $\times$  & $\times$ &  &   \\
\hline
\hline
$\widetilde{\textrm{O}5}$/$\widetilde{\textrm{D}5}^{(7)}$ & $\times$ & $\times$ & $\times$  &  & $\times$ &  & $\times$  &  & $\times$ &   \\
\hline
\end{tabular}
\caption{Spacetime-filling brane sources behind the AdS$_{3}$ flux vacua. These are the sources entering the tadpole cancellation conditions in (\ref{Tadpole_O5/D5_123}) and (\ref{Tadpole_O5/D5_4567}).}
\label{tab:sources}
\end{table}

\subsubsection{Source branes and tadpole cancellation conditions}
\label{sec:source_branes}

The space-time filling $\,\widetilde{\textrm{O}5}/\widetilde{\textrm{D}5}\,$ sources behind the AdS$_{3}$ flux vacua in (\ref{general_AdS3_1})-(\ref{general_AdS3_2}) can be identified by looking at the Bianchi identity of the background $\,F_{(3)}\,$ flux in (\ref{F3,F7_fluxes}). Using (\ref{Maurer-Cartan_eq}), it is direct to verify that
\begin{equation}
\label{Tadpole_O5/D5_123}
\left. dF_{(3)} \right|_{3456} = 0
\hspace{8mm} ,  \hspace{8mm} 
\left. dF_{(3)} \right|_{1256} = 0
\hspace{8mm} ,  \hspace{8mm} 
\left. dF_{(3)} \right|_{1234} = 0 \ ,
\end{equation}
so a so-called flux-induced tadpole\footnote{The name stems from the linear (tadpole) term $\,\int C_{6} \wedge dF_{3}\,$ that appears as a Lagrange multiplier in the first-order formulation of type IIB supergravity.} is not possible for the $\widetilde{\textrm{O}5}$/$\widetilde{\textrm{D}5}^{(1,2,3)}$ sources in Table~\ref{tab:sources}. Equivalently, the net charge of these objects vanishes in the compactification. However a non-zero tadpole can be induced by the fluxes for the $\widetilde{\textrm{O}5}$/$\widetilde{\textrm{D}5}^{(4,5,6,7)}$ sources. They are respectively given by
\begin{equation}
\label{Tadpole_O5/D5_4567}
\begin{array}{rclcl}
\left. dF_{(3)} \right|_{2457} & = &  \omega_{3} \, f_{35} + \omega_{1} \, f_{36} + \omega_{6} \, f_{37} &=& 2\kappa^{2} \,  T_{\textrm{D}5} \, \big(2N_{\widetilde{\textrm{O}5}^{(4)}} - N_{\widetilde{\textrm{D}5}^{(4)}} \big) \ , \\[3mm]
\left. dF_{(3)} \right|_{1467} & = &   \omega_{3} \, f_{34} + \omega_{5} \, f_{36} + \omega_{2} \, f_{37}  &=& 2\kappa^{2} \,  T_{\textrm{D}5} \, \big(2N_{\widetilde{\textrm{O}5}^{(5)}} - N_{\widetilde{\textrm{D}5}^{(5)}} \big) \ , \\[3mm]
\left. dF_{(3)} \right|_{2367} & = &  \omega_{1} \, f_{34} + \omega_{5} \, f_{35} + \omega_{4} \, f_{37} &=& 2\kappa^{2} \,  T_{\textrm{D}5} \, \big(2N_{\widetilde{\textrm{O}5}^{(6)}} - N_{\widetilde{\textrm{D}5}^{(6)}} \big) \ , \\[3mm]
- \left. dF_{(3)} \right|_{1357} & = & \omega_{2} \, f_{35} + 
\omega_{4} \, f_{36} + 
\omega_{6} \, f_{34}
&=&  2\kappa^{2} \,  T_{\textrm{D}5} \, \big(2N_{\widetilde{\textrm{O}5}^{(7)}} - N_{\widetilde{\textrm{D}5}^{(7)}} \big) \ ,
\end{array}
\end{equation}
where the minus sign in the last tadpole is due to our convention for $\,\varphi_{(4)}\,$ in (\ref{G2-inv_forms}). The above flux combinations are precisely the ones entering the last two lines in the scalar potential (\ref{V_N1}). Let us highlight at this point that neither the $\,F_{(3)}\,$ flux components $(f_{31},f_{32},f_{33})$ nor the $\,F_{(7)}\,$ flux $\,f_{7}\,$ enter the tadpole cancellation conditions in (\ref{Tadpole_O5/D5_4567}). As we just saw in (\ref{scalings_N}), these are the fluxes that control the hierarchy of scales at the AdS$_{3}$ flux vacua in (\ref{general_AdS3_1})-(\ref{general_AdS3_2}). In the next section we will trace back these unrestricted fluxes to obtain partial information about the branes behind the AdS$_{3}$ flux vacua following the prescription of \cite{Apers:2025pon,Apers:2026lgi}.

It is also worth emphasising that the Lie brackets in (\ref{algebra_internal_space}) specify a proper seven-dimensional Lie algebra. As such, the associated structure constants $\,\omega_{mn}{}^{p}\,$ ($m=1,\ldots,7$) obey the Jacobi identity 
\begin{equation}
\label{Jacobi_identity}
\omega_{[mn}{}^{q} \, \omega_{p]q}{}^{r} = 0 \ .
\end{equation}
In \cite{Villadoro:2007yq}, the inclusion of KK-monopoles as sources in an AdS flux vacuum was connected to the violation of the Jacobi identity in (\ref{Jacobi_identity}), thus extending beyond ordinary Scherk--Schwarz (twisted torus) reductions \cite{Scherk:1979zr}. Examples of AdS$_{4}$ flux vacua in type IIA/M-theory with (smeared) KK-monopoles as sources were studied in \cite{Derendinger:2014wwa,Danielsson:2014ria}. Remarkably, it was shown there that, in contrast to D-branes/M-branes, KK-monopole sources are effectively encoded within the bulk action of eleven-dimensional or type II supergravity.

\begin{table}[t]
\centering
\renewcommand{\arraystretch}{1.5}
\begin{tabular}{|c||cc||c||cc|cc|cc||c||c|}
\hline
 & $dt$ & $dx$ & $dy$ & $\eta^{1}$  & $\eta^{2}$ & $\eta^{3}$  & $\eta^{4}$ & $\eta^{5}$ & $\eta^{6}$ & $\eta^{7}$ & Flux \\ 
\hline
\hline
\textrm{D1} & $\times$ & $\times$ & & & & & & & & & $f_{7}$ \\
\hline
\hline
$\textrm{O}5$/$\textrm{D}5^{(1)}$   & $\times$ & $\times$ &  &  &   & $\times$ & $\times$ & $\times$ & $\times$ & & $f_{31}$ \\ 
$\textrm{O}5$/$\textrm{D}5^{(2)}$ & $\times$ & $\times$ &  &  $\times$ & $\times$ &  &   & $\times$ & $\times$ &  & $f_{32}$ \\
$\textrm{O}5$/$\textrm{D}5^{(3)}$   & $\times$ & $\times$ &  & $\times$ & $\times$  & $\times$ & $\times$ &   &   & & $f_{33}$ \\
\hline
\hline
$\textrm{O}5$/$\textrm{D}5^{(4)}$ & $\times$ & $\times$ &  &  & $\times$ &  & $\times$ & $\times$ &   & $\times$ & $f_{34}$ \\
$\textrm{O}5$/$\textrm{D}5^{(5)}$ & $\times$ & $\times$ &  & $\times$ &   &  & $\times$ &  & $\times$ & $\times$ & $f_{35}$\\ 
$\textrm{O}5$/$\textrm{D}5^{(6)}$ & $\times$ & $\times$ &  &  & $\times$ & $\times$ &   &  & $\times$ &  $\times$ & $f_{36}$  \\
\hline
\hline
$\textrm{O}5$/$\textrm{D}5^{(7)}$ & $\times$ & $\times$ &  & $\times$ &  & $\times$ &   & $\times$ &  & $\times$ & $f_{37}$  \\
\hline
\hline
 & $\times$ & $\times$ &  & $\bullet$ &  & $\times$ & $\times$ & $\times$ & $\times$  &  &  $\omega_{1}$ \\
 KK5$^{(1,a)}$ & $\times$ & $\times$ &  & $\times$ & $\times$& $\bullet$ &   & $\times$ & $\times$  &  & $\omega_{3}$  \\
 & $\times$ & $\times$ &  & $\times$ & $\times$ & $\times$ & $\times$ & $\bullet$ & &  &  $\omega_{5}$ \\
\hline
\hline
 & $\times$ & $\times$ &  &  & $\bullet$ & $\times$ & $\times$ & $\times$ & $\times$  &  &  $\omega_{2}$ \\
 KK5$^{(2,i)}$ & $\times$ & $\times$ &  & $\times$ & $\times$&  & $\bullet$   & $\times$ & $\times$  &  & $\omega_{4}$  \\
 & $\times$ & $\times$ &  & $\times$ & $\times$ & $\times$ & $\times$ & & $\bullet$ &  & $\omega_{6}$  \\
\hline
\end{tabular}
\caption{Domain-wall branes and KK5 monopoles responsible for the fluxes supporting the AdS$_{3}$ flux vacua that emerge in the near-horizon region of the brane intersection. Observe that there is only one overall transverse direction, $\,y$, which is simultaneously transverse to all the domain-wall branes and KK5 monopoles.}
\label{tab:DW-branes}
\end{table}

\subsubsection{Domain-wall branes}
\label{sec:DW_branes}

The constant fluxes supporting an AdS$_{3}$ flux vacuum originate from a codimension-one domain-wall brane intersection whose near-horizon region describes precisely such an AdS$_{3}$ flux background. All the domain-wall branes, amongst which we have to include KK$5$ monopoles to account for the metric fluxes in (\ref{Maurer-Cartan_eq}), must share one direction, we denote it $\,y$, as an overall transverse direction for the brane intersection to depend on it. The domain-wall branes and KK5-monopoles responsible for the gauge and metric fluxes in (\ref{F3,F7_fluxes}) and (\ref{Maurer-Cartan_eq}), which support the AdS$_{3}$ flux vacua in (\ref{general_AdS3_1})-(\ref{general_AdS3_2}), are summarised in Table~\ref{tab:DW-branes}.

Due to the generic $\,y$-dependence of the domain-wall brane intersection one has non-zero components of the Bianchi identity of $\,F_{(3)}\,$ of the form
\begin{equation}
\label{dF3_5-branes}
\left. dF_{(3)} \right|_{yai7} \neq 0
\hspace{8mm} , \hspace{8mm} 
\left. dF_{(3)} \right|_{yibc} \neq 0
\hspace{8mm} , \hspace{8mm}
\left. dF_{(3)} \right|_{yijk} \neq 0 \ .
\end{equation}
These add up to $3+3+1$ non-zero components, and are in one-to-one correspondence with the seven types of O$5$/D$5$ domain-wall branes displayed in Table~\ref{tab:DW-branes}. Together with the O$5$/D$5$, there are also the O$1$/D$1$ domain-wall branes for which the relevant Bianchi identity reads
\begin{equation}
\label{dF7_1-branes}
\left. dF_{(7)} \right|_{yaibjck7} \neq 0 \ .  
\end{equation}
However, both (\ref{dF3_5-branes}) and (\ref{dF7_1-branes}) vanish in the near-horizon region where the AdS$_{3}$ flux vacua are recovered. This is so because the flux parameters become \textit{constant} at the horizon region, and also because there are no metric flux components of the form $\,\omega_{ym}{}^{n}\,$ with $\,m=1, \ldots, 7$. Therefore all the domain-wall D1/D5-branes in Table~\ref{tab:DW-branes} are replaced by the corresponding constant flux at the AdS$_{3}$ flux vacua.

The situation for the domain-wall KK5-monopoles is more subtle since they are intrinsic to the geometry. In this work, we restrict our analysis to domain-wall KK5-monopoles for which the metric fluxes in (\ref{Maurer-Cartan_eq}) are independent of the transverse coordinate $\,y$, namely,
\begin{equation}
\label{constant_flux_cond}
d\omega_{a}=d\omega_{i}=0 \ .
\end{equation}
This has two consequences: $i)$ The twisted nature of the internal space remains the same all over the ten-dimensional ($y$-dependent) domain-wall solution. $ii)$ Unlike for the domain-wall D5/D1-branes entering the right hand side of (\ref{dF3_5-branes}) and (\ref{dF7_1-branes}), no domain-wall KK5-monopoles have to be considered when solving the ten-dimensional equations of motion of type IIB supergravity (see Appendix~\ref{app:10D_EOMs} for details).

\section{Flux backtracking AdS$_{3}$ flux vacua}
\label{sec:flux_backtracking}

In this section we will gain some insight into the brane intersection underlying the AdS$_{3}$ flux vacua in (\ref{general_AdS3_1})-(\ref{general_AdS3_2}) following the method introduced in \cite{Apers:2025pon,Apers:2026lgi}. The method exploits both the existence of a consistent $D=3$ supergravity capturing the AdS$_3$ flux vacua and explicit uplift formulae that embed any three-dimensional solution into ten-dimensional type~IIB supergravity.

The starting point is the $\,\mathcal{N}=1$, $\,D=3\,$ superpotential in (\ref{W_model}), which includes the metric fluxes in (\ref{Maurer-Cartan_eq}) and the gauge fluxes in (\ref{F3,F7_fluxes}). To make the discussion concrete, let us consider a generic AdS$_3$ flux vacuum of the form (\ref{general_AdS3_1})-(\ref{general_AdS3_2}), in which \emph{all} gauge fluxes are switched on. If a subset of these gauge fluxes is subsequently turned off, the resulting superpotential generally no longer admits the same AdS$_3$ flux vacuum as a critical point. Nevertheless, when the deactivated fluxes are precisely those that remain unconstrained in the AdS$_3$ flux vacua, in our case,
\begin{equation}
\label{fluxes_unrestricted}
f_{7}
\hspace{10mm} \textrm{ and } \hspace{10mm}
\left(f_{31}  \,,\, f_{32}  \,,\,  f_{33} \right) \ ,
\end{equation}
the resulting supergravity theory still accommodates three-dimensional domain-wall (DW$_3$) solutions. The Ansatz for DW$_{3}$ solutions with a flat slicing is of the form
\begin{equation}
\label{DW_ansatz}
ds_{3}^{2} = dy^{2} + e^{2 A(y)} \, ds_{2}^{2}
\hspace{10mm} \textrm{ and } \hspace{10mm}
\phi^{A}=\phi^{A}(y) \ ,
\end{equation}
with $\,y \in (0,\infty)\,$ and $\,ds_{2}^{2}=-dt^{2} + dx^{2}$. Plugging this Ansatz into the action (\ref{S_3D}), and completing squares \`a la Bogomol'nyi–Prasad–Sommerfield (BPS), one arrives at a set of BPS equations of the form \cite{Skenderis:1999mm}
\begin{equation}
\label{BPS_eqs}
{\phi'}^{A} = \mp K^{AB} \, \frac{\partial W}{\partial \phi^B}, 
\hspace{10mm} \textrm{ and } \hspace{10mm}
A' = \pm W   \ ,
\end{equation}
where primes denote derivatives with respect to the coordinate $\,y$. Uplifting the DW$_{3}$ solutions back to ten dimensions using (\ref{10D_metric})-(\ref{ds7_metric}) and (\ref{gs&vol7}) produces a solution of type IIB supergravity that we will refer to as the \textit{background solution}. The name reflects the fact that it describes the type IIB background probed by the D$1$-branes and the D$5^{(1,2,3)}$-branes in Table~\ref{tab:DW-branes}, which are responsible for the unrestricted fluxes in (\ref{fluxes_unrestricted}).

\subsection{The 3D domain-wall solutions}

Let us set to zero the fluxes in (\ref{fluxes_unrestricted}) which, as recalled above, are unrestricted and do not enter the tadpole cancellation conditions in (\ref{Tadpole_O5/D5_4567}). Setting 
\begin{equation}
\label{Flux_DW}
f_{7}=0
\hspace{10mm} \textrm{ and }  \hspace{10mm} 
f_{31}=f_{32}=f_{33}=0 \ ,
\end{equation}
the system of BPS equations in (\ref{BPS_eqs}) possesses DW$_{3}$ solutions of the form (\ref{DW_ansatz}) with
\begin{equation}
\label{DW_sol_general}
\tau(y) = c_{\tau} \, y^{\frac{3}{4}}
\hspace{6mm} , \hspace{6mm} 
\rho(y) = c_{\rho} \, y^{\frac{1}{28}}
\hspace{6mm} , \hspace{6mm} 
\ell_{a}(y) = c_{\ell_{a}} \, y^{-\frac{1}{28}}
\hspace{6mm} , \hspace{6mm} 
\ell_{i}(y) = c_{\ell_{i}} \, y^{-\frac{1}{28}} \ ,
\end{equation}
where the coefficients $\,(c_{\tau},c_{\ell_{a}},c_{\ell_{i}})\,$ depend on the metric fluxes $\,(\omega_{a},\omega_{i})$, as well as on those gauge fluxes $(f_{34},f_{35},f_{36},f_{37})$ which were not set to zero. Since a given AdS$_3$ flux vacuum corresponds to a specific choice of fluxes in (\ref{general_AdS3_1})--(\ref{general_AdS3_2}), the coefficients $\,(c_{\tau},c_{\ell_{a}},c_{\ell_{i}})\,$ generally take different values for different AdS$_3$ flux vacua. On the contrary, the coefficient $\,c_{\rho}\,$ is arbitrary in the DW$_{3}$ solutions and the warp factor is always given by
\begin{equation}
\label{DW_e2A_general}
e^{2A(y)} = y^{\frac{5}{4}} \ .  
\end{equation}
Some examples of DW$_{3}$ solutions associated with different AdS$_{3}$ flux vacua are presented in Appendix~\ref{app:AdS3_vacua}. Nevertheless, as we will show now, all the DW$_{3}$ solutions of the form (\ref{DW_sol_general})-(\ref{DW_e2A_general}) describe the same class of type IIB backgrounds to be probed by the D$1$- and D$5^{(1,2,3)}$-branes.

\subsection{Uplifting the 3D domain-walls: the type IIB background solution}
\label{sec:background_solution}

Uplifting the DW$_{3}$ solutions in (\ref{DW_sol_general})-(\ref{DW_e2A_general}) to ten-dimensions using (\ref{10D_metric})-(\ref{ds7_metric}) and (\ref{gs&vol7}), one obtains the following class of type IIB backgrounds. The ten-dimensional geometry and dilaton are given by
\begin{equation}
\label{10D_metric_background}
\begin{array}{rcl}
ds_{10}^2 & = &    \hat{y}^{-1} \, d\hat{s}_{2}^2 + \left[  d\hat{y}^2 + \hat{y}^{2} \,  \left(\hat{\eta}^{7}\right)^{2}  \right] 
+  \left[ \, \left(\hat{\eta}^{a}\right)^{2} + \left( \dfrac{c_{\ell_{i}}}{c_{\ell_{a}}} \right)^{2} \left(\hat{\eta}^{i}\right)^{2} \right] \ , \\[2mm]
g_{s}^{2} 
\,\,\,=\,\,\, e^{2\Phi} 
&=& \hat{y}^{-2} \ ,
\end{array}
\end{equation}
where we have first introduced a set of new coordinates and one-form basis elements\footnote{The rescaling of the one-form basis in (\ref{c.o.c_background_solution}) is compatible with a rescaling of the $x$-coordinates in the generic case of (\ref{one-form_integrated_general}) accompanied by a suitable (inverse) rescaling of the metric fluxes $\,(\omega_{a},\omega_{i})$.}
\begin{equation}
\label{c.o.c_background_solution}
y = \left(\frac{c_{\tau}}{4}\right)^{4}  \, \hat{y}^{4}
\hspace{3mm} , \hspace{3mm}
\left(t,x\right) =\left(\frac{c_{\tau}^3}{4}\right)^{\frac{1}{2}}\left(\hat{t},\hat{x} \right)
\hspace{3mm} , \hspace{3mm}
(\eta^{a},\eta^{i}) = p^{-1} \, (\hat{\eta}^{a},\hat{\eta}^{i})
\hspace{3mm} , \hspace{3mm}
\eta^{7} = q^{-1} \, \hat{\eta}^{7} \ ,
\end{equation}
with
\begin{equation}
p \equiv c_{\rho} \, c_{\ell_{a}}
\hspace{10mm} \textrm{ and } \hspace{10mm}
q  \equiv  \dfrac{c_{\rho \, }c_{\tau}}{4 \, c_{\ell_{a}}^{3} c_{\ell_{i}}^{3}}
\ ,
\end{equation}
and then we have fixed the arbitrary $\,c_{\rho}\,$ coefficient as $\,c_{\rho}^{7} = \tfrac{c_{\tau}^{3}}{16}\,$. The background gauge fluxes are of the form 
\begin{equation}
\label{F3_background}
\begin{array}{rcl}
F_{(3)} &=& - f_{34} \, \eta^{136} - f_{35} \, \eta^{235} -  f_{36} \, \eta^{145}   +  f_{37} \, \eta^{246}  \ , \\[2mm]
F_{(7)} &=& 0 \ .
\end{array}
\end{equation}
We have verified that the above type IIB background solutions satisfy the ten-dimensional equations of motion in Appendix~\ref{app:10D_EOMs}. The string coupling in (\ref{10D_metric_background}) diverges at $\,\hat{y}=0\,$ where a strongly-coupled curvature singularity appears. As observed in \cite{Apers:2025pon} for the DGKT flux vacua \cite{DeWolfe:2005uu}, this singularity does not appear to admit a UV completion in terms of weakly-coupled D-branes and open strings, so a standard brane-engineering derivation of the dual CFT seems unplausible at this point.

\subsection{Reinstating D1- and D5$^{(1,2,3)}$-branes: the 10D interpolating solution}
\label{sec:interpolating_solution}

Let us consider the background solution in (\ref{10D_metric_background})-(\ref{F3_background}) and introduce a set of harmonic-like $H$-functions associated with the D$1$-branes and D$5^{(1,2,3)}$-branes probing it. In other words, let us ``zoom out'' the branes responsible for the unrestricted fluxes in (\ref{fluxes_unrestricted}). Following the harmonic superposition principle, the ten-dimensional metric and dilaton take the form
\begin{equation}
\label{metric_flux_backtrack}
\begin{array}{rcl}
ds_{10}^2 & = &  \left(H_1 \, H_{5}^{(1)} \, H_{5}^{(2)} \, H_{5}^{(3)} \right)^{-\frac{1}{2}}  \, \hat{y}^{-1} \,\, d\hat{s}_{2}^2 
+ \left(H_1 \, H_{5}^{(1)} \, H_{5}^{(2)} \, H_{5}^{(3)} \right)^{\frac{1}{2}}
\left[  d\hat{y}^2  + \hat{y}^{2} \, \left(\hat{\eta}^{7}\right)^{2}
\right]
\\[6mm]
&+& \left(\dfrac{H_1 \, H_{5}^{(1)}}{H_{5}^{(2)} \, H_{5}^{(3)}}\right)^{\frac{1}{2}} \, \left[  \left(\hat{\eta}^{1}\right)^{2} + \left( \dfrac{c_{\ell_{2}}}{c_{\ell_{1}}} \right)^{2}  \left(\hat{\eta}^{2}\right)^{2}  \right]  \\[6mm]
&+& \left(\dfrac{H_1 \, H_{5}^{(2)}}{H_{5}^{(3)} \, H_{5}^{(1)}}\right)^{\frac{1}{2}} \,  \left[  \left(\hat{\eta}^{3}\right)^{2} + \left( \dfrac{c_{\ell_{4}}}{c_{\ell_{3}}} \right)^{2}  \left(\hat{\eta}^{4}\right)^{2}  \right]  \\[6mm]
&+& \left(\dfrac{H_1 \, H_{5}^{(3)}}{H_{5}^{(1)} \, H_{5}^{(2)}}\right)^{\frac{1}{2}}           \,  \left[ \left(\hat{\eta}^{5}\right)^{2} + \left( \dfrac{c_{\ell_{6}}}{c_{\ell_{5}}} \right)^{2}  \left(\hat{\eta}^{6}\right)^{2}  \right] \ ,
\end{array}
\end{equation}
and
\begin{equation}
\label{dilaton_flux_backtrack}
\begin{array}{rcl}
e^{2\Phi} & = & \left(\dfrac{H_{1}}{H_{5}^{(1)} \, H_{5}^{(2)} \, H_{5}^{(3)}}\right) \, \hat{y}^{-2}  \ .
\end{array}
\end{equation}
The $H$-functions for the D$1$-branes and D$5^{(1,2,3)}$-branes that we are reinstating in the background solution (\ref{10D_metric_background})-(\ref{F3_background}) are found to be
\begin{equation}
\label{harmonic_function_D1_D5_yhat}
H_{1} = 1 - \dfrac{q_{1}}{\hat{y}} 
\hspace{6mm} , \hspace{6mm}
H_{5}^{(1)} = 1 - \dfrac{q_{5,1}}{\hat{y}} 
\hspace{6mm} , \hspace{6mm}
H_{5}^{(2)} = 1 - \dfrac{q_{5,2}}{\hat{y}} 
\hspace{6mm} , \hspace{6mm}
H_{5}^{(3)} = 1 - \dfrac{q_{5,3}}{\hat{y}} \ ,
\end{equation}
with D$1$- and D$5^{(1,2,3)}$-brane charges given by
\begin{equation}
\label{q1_charge_interpolating}
q_{1} = \left( \frac{2^{6}}{c_{\tau}^{4}} \right) \, f_{7}  \ ,
\end{equation}
and
\begin{equation}
\label{q5_charges_interpolating}
q_{5,1} = \left( \dfrac{2^{\frac{26}{7}} \, c_{\ell_{a}}^{2} \, c_{\ell_{i}}^{2}}{c_{\tau}^{\frac{16}{7}}} \right) \, f_{31} 
\hspace{5mm} , \hspace{5mm}
q_{5,2} =\left( \dfrac{2^{\frac{26}{7}} \, c_{\ell_{a}}^{2} \, c_{\ell_{i}}^{2}}{c_{\tau}^{\frac{16}{7}}} \right) \, f_{32}
\hspace{5mm} , \hspace{5mm}
q_{5,3} = \left( \dfrac{2^{\frac{26}{7}} \, c_{\ell_{a}}^{2} \, c_{\ell_{i}}^{2}}{c_{\tau}^{\frac{16}{7}}} \right) \, f_{33} \ ,
\end{equation}
so that $\,q_{1} \,,\, q_{5,1} \,,\, q_{5,2} \,,\, q_{5,3} < 0\,$ by virtue of (\ref{fluxes_negative}). On the other hand, the RR three-form field strength can be expressed as
\begin{equation}
\label{F3=F3ele+F3mag}
F_{(3)} = F_{(3)}^{(ele)} + F_{(3)}^{(mag)} \ ,
\end{equation}
with components given by
\begin{equation}
\label{F3_flux_backtrack}
\begin{array}{rcl}
F_{(3)}^{(ele)} &=& d\left(H_{1}^{-1}\right) \,  d\hat{t} \wedge d\hat{x} \ , \\[6mm]
F_{(3)}^{(mag)} &=& f_{31} \, \eta^{127}  +  f_{32} \, \eta^{347} + f_{33} \, \eta^{567}  - f_{34} \, \eta^{136} - f_{35} \, \eta^{235} -  f_{36} \, \eta^{145}   +  f_{37} \, \eta^{246} \ , 
\end{array}
\end{equation}
so that $\,F_{(3)}^{(mag)}\,$ recovers (\ref{F3,F7_fluxes}) and
\begin{equation}
\label{F7_flux_backtrack}
F_{(7)} = - \star F_{(3)}^{(ele)} = f_{7} \, \eta^{1234567} \ .   
\end{equation}
We have again verified that the solution in (\ref{metric_flux_backtrack})-(\ref{F7_flux_backtrack}) satisfies the ten-dimensional equations of motion in Appendix~\ref{app:10D_EOMs}. The key feature of this solution is that it interpolates between the type IIB background solution (\ref{10D_metric_background})-(\ref{F3_background}) in the asymptotic region at $\,\hat{y}\rightarrow\infty\,$ (where $\,H_{1}=H_{5}^{(1)}=H_{5}^{(2)}=H_{5}^{(3)}=1$), and the AdS$_3$ flux vacua in (\ref{general_AdS3_1})--(\ref{general_AdS3_2}) in the near-horizon region at $\,\hat{y}\rightarrow 0$. Let us now examine this near-horizon region in more detail.

\subsubsection*{Near-horizon region}

By taking the near-horizon limit, \textit{i.e.} $\hat{y}\rightarrow 0$, of the $H$-functions in (\ref{harmonic_function_D1_D5_yhat}), the above type IIB interpolating solution becomes
\begin{equation}
\label{NH-region_metric}
\begin{array}{rcl}
ds_{10}^2 & = & \left(q_1 \, q_{5,1} \, q_{5,2} \, q_{5,3} \right)^{\frac{1}{2}} \left[ \left(q_1 \, q_{5,1} \, q_{5,2} \, q_{5,3} \right)^{-1}  \, \hat{y} \,\, d\hat{s}_{2}^2 + \dfrac{d\hat{y}^2}{\hat{y}^{2}} \right] \\[6mm]
&+& \left(\dfrac{q_1 \, q_{5,1}}{q_{5,2} \, q_{5,3}}\right)^{\frac{1}{2}} \, \left[  \left(\hat{\eta}^{1}\right)^{2} + \left( \dfrac{c_{\ell_{2}}}{c_{\ell_{1}}} \right)^{2} \, \left(\hat{\eta}^{2}\right)^{2}  \right]  \\[6mm]
&+& \left(\dfrac{q_1 \, q_{5,2}}{q_{5,3} \, q_{5,1}}\right)^{\frac{1}{2}} \,  \left[  \left(\hat{\eta}^{3}\right)^{2} + \left( \dfrac{c_{\ell_{4}}}{c_{\ell_{3}}} \right)^{2} \, \left(\hat{\eta}^{4}\right)^{2}  \right]  \\[6mm]
&+& \left(\dfrac{q_1 \, q_{5,3}}{q_{5,1} \, q_{5,2}}\right)^{\frac{1}{2}}           \,  \left[ \left(\hat{\eta}^{5}\right)^{2} + \left( \dfrac{c_{\ell_{6}}}{c_{\ell_{5}}} \right)^{2} \, \left(\hat{\eta}^{6}\right)^{2}  \right] \\[6mm]
&+&  \left(q_1 \, q_{5,1} \, q_{5,2} \, q_{5,3} \right)^{\frac{1}{2}} \left(\hat{\eta}^{7}\right)^{2} \ ,
\end{array}
\end{equation}
and
\begin{equation}
\label{NH-region_dilaton}
\begin{array}{rcl}
e^{2\Phi} & = & \left(\dfrac{q_{1}}{q_{5,1} \, q_{5,2} \, q_{5,3}}\right)  \ ,
\end{array}
\end{equation}
with the D$1$- and D$5^{(1,2,3)}$-brane charges given in (\ref{q1_charge_interpolating}) and (\ref{q5_charges_interpolating}). The $\,F_{(3)}\,$ field strength in (\ref{F3=F3ele+F3mag}) has components
\begin{equation}
\label{NH-region_F3}
\begin{array}{rcl}
F_{(3)}^{(ele)} &=& - q_{1}^{-1} \,\, d\hat{y} \wedge d\hat{t} \wedge d\hat{x} \ , \\[6mm]
F_{(3)}^{(mag)} &=& f_{31} \, \eta^{127}  +  f_{32} \, \eta^{347} + f_{33} \, \eta^{567}  - f_{34} \, \eta^{136} - f_{35} \, \eta^{235} -  f_{36} \, \eta^{145}   +  f_{37} \, \eta^{246} \ , 
\end{array}
\end{equation}
so that
\begin{equation}
\label{NH-region_F7}
F_{(7)} = - \star F_{(3)}^{(ele)} = f_{7} \, \eta^{1234567} \ .   
\end{equation}

The near-horizon geometry in (\ref{NH-region_metric}) describes an $\,\textrm{AdS}_{3} \times \mathcal{M}_{7}\,$ flux vacuum with $\,\mathcal{M}_{7}\,$ being associated (after quotient) with the seven-dimensional solvmanifold in (\ref{ds7_metric}) with one-form basis elements (\ref{one-form_integrated_general}). The values of the internal radii in (\ref{NH-region_metric}) and the type IIB dilaton in (\ref{NH-region_dilaton}) perfectly agree with the moduli VEV's at the AdS$_{3}$ flux vacua obtained from (\ref{general_AdS3_1})-(\ref{general_AdS3_2}). The gauge fluxes in (\ref{NH-region_F3}) and (\ref{NH-region_F7}) correctly reproduce the ones in (\ref{F3,F7_fluxes}) used in the flux compactification. As a last sanity check, the compatibility between the $\,\textrm{AdS}_{3}\,$ radius, as read off from the near-horizon metric (\ref{NH-region_metric}), and the (string frame) metric Ansatz in (\ref{10D_metric}), forces the relation\footnote{The factor of $\,4\,$ in the left hand side of (\ref{consistency_AdS_radius}) is due to the non-standard form of the AdS$_{3}$ factor in (\ref{NH-region_metric}).}
\begin{equation}
\label{consistency_AdS_radius}
4 \left(q_1 \, q_{5,1} \, q_{5,2} \, q_{5,3} \right)^{\frac{1}{2}} = \tau^{-2} L^2 \ ,
\end{equation}
where $\,L^2=-\frac{2}{V_{0}}\,$ and $\,V_{0}\,$ is the vacuum energy at the corresponding AdS$_{3}$ extremum of the scalar potential in (\ref{V_N1}). We have verified that, indeed, (\ref{consistency_AdS_radius}) holds in all the examples of AdS$_{3}$ flux vacua collected in Appendix~\ref{app:AdS3_vacua}.

\section{The full D1-D5-KK5 intersection}
\label{sec:full_intersection}

Following the approach of \cite{Kounnas:2007dd}, we now take one step further and also ``zoom out'' the D$5^{(4,5,6,7)}$-branes. This procedure yields the complete D1-D5-KK5 intersection whose near-horizon region reproduces the $\,\mathrm{AdS}_3\times\mathcal{M}_7\,$ flux vacua. By applying the harmonic superposition principle \cite{Tseytlin:1996bh}, the ten-dimensional spacetime metric associated with the D1-D5-KK5 intersection of Table~\ref{tab:DW-branes} can be written in the following lengthy (yet remarkably symmetric) form
\begin{equation}
\label{general_domain_wall}
\begin{array}{rcl}
ds_{10}^2 & = &  \left(H_1 \, H_{5}^{(1)} \, H_{5}^{(2)} \, H_{5}^{(3)} \right)^{-\frac{1}{2}}\left(H_{5}^{(4)} \, H_{5}^{(5)} \, H_{5}^{(6)}\, H_{5}^{(7)} \right)^{-\frac{1}{2}} \, ds_{2}^2 \\[6mm]
&+& \left(H_1 \, H_{5}^{(1)} \, H_{5}^{(2)} \, H_{5}^{(3)} \right)^{\frac{1}{2}}\left(H_{5}^{(4)} \, H_{5}^{(5)} \, H_{5}^{(6)}\, H_{5}^{(7)} \right)^{\frac{1}{2}} \, \left(\displaystyle\prod_a H_{\textrm{KK}}^{(a)}\right)\,\left(\displaystyle\prod_i H_{\textrm{KK}}^{(i)}\right)  \,  dR^2 
\\[6mm]
&+&  \left(H_1 \, H_{5}^{(1)} \, H_{5}^{(2)} \, H_{5}^{(3)} \right)^{\frac{1}{2}}\left(H_{5}^{(4)} \, H_{5}^{(5)} \, H_{5}^{(6)}\, H_{5}^{(7)} \right)^{-\frac{1}{2}}  \,  \left(\displaystyle\prod_a H_{\textrm{KK}}^{(a)}\right)\,\left(\displaystyle\prod_i H_{\textrm{KK}}^{(i)}\right) \, \left(\eta^{7}\right)^{2} \\[6mm]
&+& \left(\dfrac{H_1 \, H_{5}^{(1)}}{H_{5}^{(2)}\,H_{5}^{(3)}}\right)^{\frac{1}{2}} 
\,
\left[ \left(\dfrac{H_{5}^{(4)}\, H_{5}^{(6)}}{H_{5}^{(5)} \, H_{5}^{(7)}}\right)^{\frac{1}{2}}
\left(\dfrac{H_{\textrm{KK}}^{(2)}}{H_{\textrm{KK}}^{(1)}} \right) \, \left(\eta^{1}\right)^{2} 
+
\left(\dfrac{H_{5}^{(5)}\, H_{5}^{(7)}}{H_{5}^{(4)}\,H_{5}^{(6)}}\right)^{\frac{1}{2}}
\,
\left(\dfrac{H_{\textrm{KK}}^{(1)}}{H_{\textrm{KK}}^{(2)}} \right)
\, \left(\eta^{2}\right)^{2} \right] \\[6mm]
&+& \left(\dfrac{H_1 \, H_{5}^{(2)}}{H_{5}^{(3)}\,H_{5}^{(1)}}\right)^{\frac{1}{2}}
\,
\left[
\left(\dfrac{H_{5}^{(4)}\, H_{5}^{(5)}}{H_{5}^{(6)} \, H_{5}^{(7)}}\right)^{\frac{1}{2}}
\,
\left(\dfrac{H_{\textrm{KK}}^{(4)}}{H_{\textrm{KK}}^{(3)}} \right) \, \left(\eta^{3}\right)^{2}
+ \left(\dfrac{H_{5}^{(6)} \, H_{5}^{(7)}}{H_{5}^{(4)}\,H_{5}^{(5)}}\right)^{\frac{1}{2}}
\,
\left(\dfrac{H_{\textrm{KK}}^{(3)}}{H_{\textrm{KK}}^{(4)}}\right)
\left(\eta^{4}\right)^{2} \right] \\[6mm]
&+& \left(\dfrac{H_1 \,H_{5}^{(3)}}{H_{5}^{(1)}\,H_{5}^{(2)}}\right)^{\frac{1}{2}}
\,
\left[  \left(\dfrac{H_{5}^{(5)}\, H_{5}^{(6)}}{H_{5}^{(4)}\, H_{5}^{(7)}}\right)^{\frac{1}{2}}
\,
\left(\dfrac{H_{\textrm{KK}}^{(6)}}{H_{\textrm{KK}}^{(5)}}\right) 
\, \left(\eta^{5}\right)^{2}  
+
\left(\dfrac{H_{5}^{(4)}\,H_{5}^{(7)}}{H_{5}^{(5)}\,H_{5}^{(6)}}\right)^{\frac{1}{2}}
\,
\left(\dfrac{H_{\textrm{KK}}^{(5)}}{H_{\textrm{KK}}^{(6)}}\right)
\,\left(\eta^{6}\right)^{2} \right] 
\end{array}
\end{equation}
where we have introduced harmonic-like $H$-functions accounting for \textit{all} the domain-wall branes and KK$5$-monopoles in Table~\ref{tab:DW-branes}. The KK$5$-monopoles are responsible for the \textit{constant} metric fluxes $\,(\omega_{a},\omega_{i})\,$ entering the structure equations (\ref{Maurer-Cartan_eq}) and are obtained as
\begin{equation}
\label{omega_interpolating_solution_D1-D5}
\begin{array}{lclclcrl}
\omega_1\,=\,-\dfrac{\left(H_{\textrm{KK}}^{(1)}\right)'}{H_\textrm{KK}^{(2)}\,H_{5}^{(4)}\,H_{5}^{(6)}} & \,\,\, , \,\,\, &  \omega_3\,=\,-\dfrac{\left(H_\textrm{KK}^{(3)}\right)'}{H_\textrm{KK}^{(4)}\,H_{5}^{(4)}\,H_{5}^{(5)}} & \,\,\, , \,\,\, &  \omega_5\,=\,-\dfrac{\left(H_\textrm{KK}^{(5)}\right)'}{H_\textrm{KK}^{(6)}\,H_{5}^{(5)}\,H_{5}^{(6)}} & , \\[8mm]
\omega_2\,=\,-\dfrac{\left(H_\textrm{KK}^{(2)}\right)'}{H_\textrm{KK}^{(1)}\,H_{5}^{(5)}\,H_{5}^{(7)}} & \,\,\, , \,\,\, &  \omega_4\,=\,-\dfrac{\left(H_\textrm{KK}^{(4)}\right)'}{H_\textrm{KK}^{(3)}\,H_{5}^{(6)}\,H_{5}^{(7)}} & \,\,\, , \,\,\, &  \omega_6\,=\,-\dfrac{\left(H_\textrm{KK}^{(6)}\right)'}{H_\textrm{KK}^{(5)}\,H_{5}^{(4)}\,H_{5}^{(7)}} & .
\end{array}
\end{equation}
The ten-dimensional dilaton solely depends on the $H$-functions of the D1- and D5-branes, and is given by
\begin{equation}
\begin{array}{rcl}
e^{2\Phi} & = & \left( \dfrac{H_{1}}{H_{5}^{(1)} \, H_{5}^{(2)} \, H_{5}^{(3)}}\right) \,  \left(H_{5}^{(4)} \, H_{5}^{(5)} \, H_{5}^{(6)} \, H_{5}^{(7)} \right)^{-1}  \ .
\end{array}
\end{equation}
In order to present the field strength $\,F_{(3)}\,$ associated with the D1-D5-KK5 intersection, it proves convenient to split it again as
\begin{equation}
F_{(3)} = F_{(3)}^{(ele)} + F_{(3)}^{(mag)} \ ,
\end{equation}
so that the pieces are given by
\begin{equation}
\label{F3_interpolating_solution_D1-D5}
\begin{array}{rcl}
F_{(3)}^{(ele)} &=&  d\left(H_{1}^{-1}\right) \wedge dt \wedge dx \ , \\[6mm]
F_{(3)}^{(mag)} &=& 
 \,-\left(H_{5}^{(1)}\right)' \, \eta^{127} 
\, - \, \left(H_{5}^{(2)}\right)' \, \eta^{347} \, - \, \left(H_{5}^{(3)}\right)' \,\eta^{567} \\[6mm]
& & +\, \dfrac{\left(H_{5}^{(4)}\right)'}{H_{\textrm{KK}}^{(1)} \,H_{\textrm{KK}}^{(3)}\,H_{\textrm{KK}}^{(6)}}\,\eta^{136} 
\, + \, \dfrac{\left(H_{5}^{(5)}\right)'}{H_{\textrm{KK}}^{(2)}\,H_{\textrm{KK}}^{(3)}\,H_{\textrm{KK}}^{(5)}}\,\eta^{235} 
\, + \,\dfrac{\left(H_{5}^{(6)}\right)'}{H_{\textrm{KK}}^{(1)}\,H_{\textrm{KK}}^{(4)}\,H_{\textrm{KK}}^{(5)}}\,\eta^{145} \\[6mm]
& & \, - \, \dfrac{\left(H_{5}^{(7)}\right)'}{H_{\textrm{KK}}^{(2)}\,H_{\textrm{KK}}^{(4)}\,H_{\textrm{KK}}^{(6)}}\, \eta^{246} \ ,
\end{array}
\end{equation}
where primes denote ordinary derivatives with respect to the overall transverse coordinate $\,R\,$ in the geometry (\ref{general_domain_wall}).

The above Ansatz for the type IIB metric, dilaton, and gauge fluxes already incorporates the fact that, in our compactification setup, no flux-induced tadpoles can arise from the D$1$- and D$5^{(1,2,3)}$-branes when reaching the horizon. As a result, these branes must remain localised, implying that the corresponding $H$-functions are harmonic in the unique overall transverse direction $\,R$. Explicitly,
\begin{equation}
\label{H-functions_color}
H_{1} = 1 - f_{7} \, R 
\hspace{4mm} , \hspace{4mm}
H_{5}^{(1)} = 1 - f_{31} \, R
\hspace{4mm} , \hspace{4mm}
H_{5}^{(2)} = 1 - f_{32} \, R
\hspace{4mm} , \hspace{4mm}
H_{5}^{(3)} = 1 - f_{33} \, R \ .
\end{equation}
The remaining freedom resides in the choice of the $\,H^{(4,5,6,7)}$-functions, since the KK5-monopole functions $\,H_{\textrm{KK}}^{(a)}\,$ and $\,H_{\textrm{KK}}^{(i)}\,$ are then uniquely determined by direct integration of \eqref{omega_interpolating_solution_D1-D5}. This brings us to our final statement: for arbitrary choices of the $\,H^{(4,5,6,7)}$-functions, the equations of motion and Bianchi identities of type IIB supergravity in the presence of the O-planes, D-branes and KK5-monopoles in Tables~\ref{tab:sources} and \ref{tab:DW-branes}, as presented in Appendix~\ref{app:10D_EOMs}, are automatically satisfied.

Importantly, the transverse coordinate $\,R\,$ in (\ref{general_domain_wall}) is not necessarily the same as the transverse coordinate $\,y\,$ (or $\,\hat{y}\,$ in (\ref{c.o.c_background_solution})) used in the previous section. Indeed, a direct comparison between (\ref{H-functions_color}) and (\ref{harmonic_function_D1_D5_yhat}) establishes the correspondence $\,R \sim \hat{y}^{-1}$. This implies that, when using the overall transverse coordinate $\,R\,$, the near-horizon region discussed in Section~\ref{sec:interpolating_solution} corresponds to the $\,R\rightarrow \infty\,$ region of (\ref{general_domain_wall}). Similarly, the asymptotic region corresponds to the $\,R\rightarrow 0\,$ region of (\ref{general_domain_wall}).

\subsection{Near-horizon and asymptotic regions}

\begin{figure}[t]
\begin{center}
\includegraphics[width=1\textwidth]{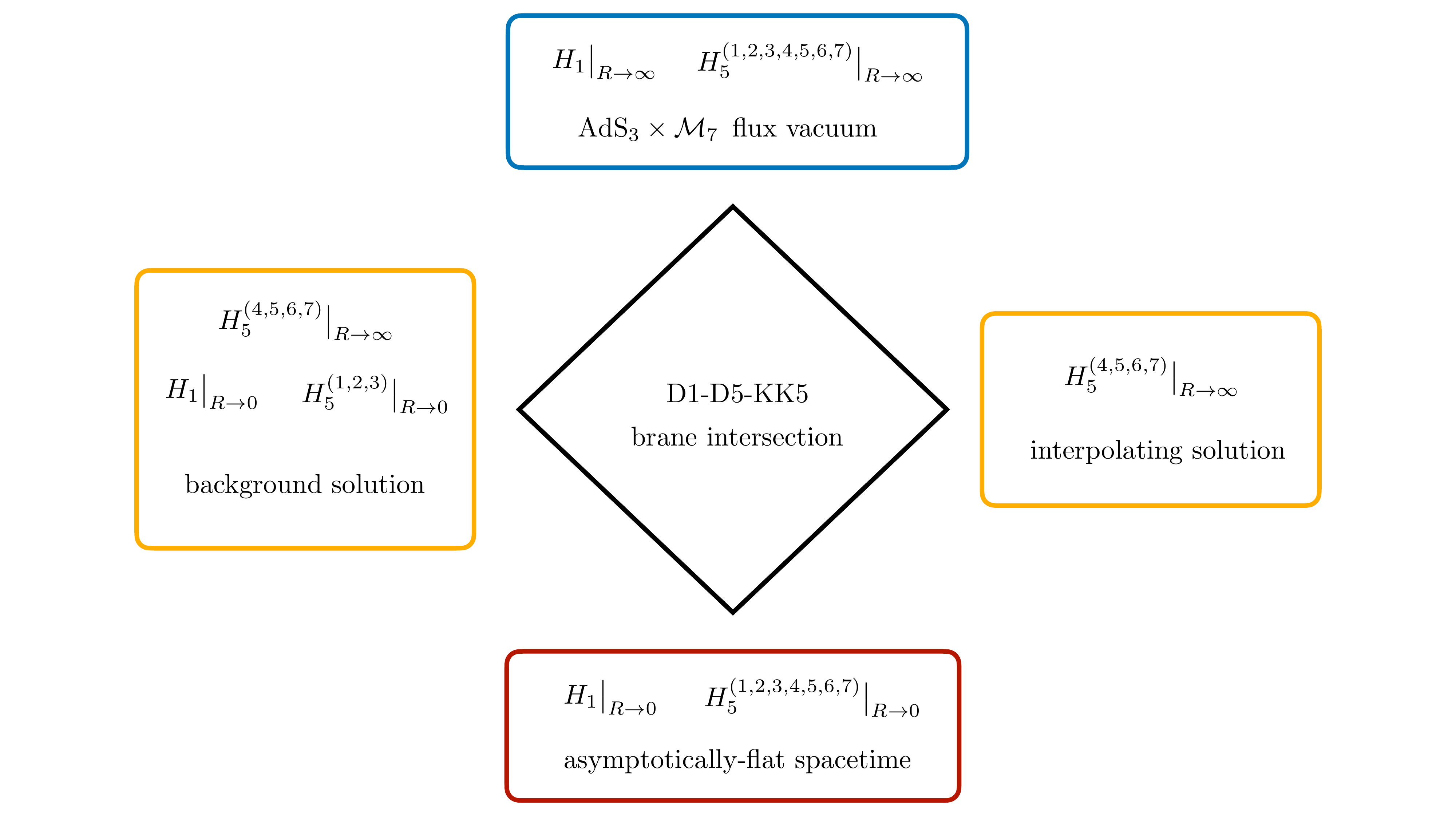} 
\end{center}
\caption{Ten-dimensional type IIB solutions discussed in Section~\ref{sec:flux_backtracking}, recovered as limits of the D1-D5-KK5 intersection.}
\label{Fig:diagram_solutions}
\end{figure}

We will show here how an $\,\textrm{AdS}_{3} \times \mathcal{M}_{7}$\, flux vacuum as well as the ten-dimensional type IIB background (see Section~\ref{sec:background_solution}) and interpolating (see Section~\ref{sec:interpolating_solution}) solutions, can be recovered from the D1-D5-KK5 intersection upon taking appropriate limits on the various $H$-functions. In addition, a genuine asymptotic limit becomes now available at $\,R \rightarrow 0\,$ in which one moves away from \textit{all} the D1- and D5-branes simultaneously recovering an asymptotically-flat spacetime. We have summarised all these limits in Figure~\ref{Fig:diagram_solutions}.

\subsubsection*{Recovering the $\,\textrm{AdS}_{3} \times \mathcal{M}_{7}\,$ flux vacuum in the full near-horizon limit}

The existence of an $\,\textrm{AdS}_{3} \times \mathcal{M}_{7}\,$ near-horizon region in the solution (\ref{general_domain_wall}) forces the $H$-functions at $\,R\rightarrow \infty\,$ to adopt a specific scaling dependence. The functions encoding the D$1$-branes and D$5^{(1,2,3)}$-branes in (\ref{H-functions_color}) are harmonic and therefore approach
\begin{equation}
\label{H1&H5_123_scalings_horizon}
H_{1} \big|_{\infty}  \sim  - f_{7} \, R 
\hspace{4mm} , \hspace{4mm}
H_{5}^{(1)} \big|_{\infty}  \sim  - f_{31} \, R
\hspace{4mm} , \hspace{4mm}
H_{5}^{(2)} \big|_{\infty}  \sim  - f_{32} \, R
\hspace{4mm} , \hspace{4mm}
H_{5}^{(3)} \big|_{\infty}  \sim  - f_{33} \,  R \ ,
\end{equation}
where the subscript in $\,|_{\infty}\,$ refers to the behaviour of the functions in the near-horizon region $\,R \rightarrow \infty$. The $H$-functions associated with the D$5^{(4,5,6,7)}$-branes must have a scaling behaviour of the form
\begin{equation}
\label{H5_4567_scalings_horizon}
\begin{array}{llll}
H_{5}^{(4)} \big|_{\infty} \sim  - \dfrac{c_{1}\,c_{3}\,c_{6}}{a_{4}} \, f_{34} \, R^{a_{4}}
& \hspace{3mm} , \hspace{3mm} &
H_{5}^{(5)}\big|_{\infty}  \sim  - \dfrac{c_{2}\,c_{3}\,c_{5}}{a_{5}} \, f_{35} \, R^{a_{5}} & , \\[6mm]
H_{5}^{(6)}\big|_{\infty}  \sim - \dfrac{c_{1}\,c_{4}\,c_{5}}{a_{6}} \, f_{36} \,  R^{a_{6}}
& \hspace{3mm} , \hspace{3mm} &
H_{5}^{(7)}\big|_{\infty}  \sim - \dfrac{c_{2}\,c_{4}\,c_{6}}{a_{7}} \,   f_{37} \, R^{a_{7}} & ,
\end{array}
\end{equation}
with
\begin{equation}
\label{simple_relation_1}
a_{4}+a_{5}+a_{6}+a_{7} = - 2 \ .
\end{equation}
The coefficients $\,(c_{a},c_{i})\,$ appear in the $H$-functions encoding the KK$^{(1,a)}$ and KK$^{(2,i)}$ monopoles. In order to recover an AdS$_{3}$ flux vacuum \textit{with constant fluxes} at $\,R \rightarrow \infty\,$, these functions must behave as
\begin{equation}
\label{HKK_scalings_horizon}
H_\textrm{KK}^{(a)}\big|_{\infty} \sim c_{a} \, R^{b_{a}}
\hspace{8mm} , \hspace{8mm} 
H_\textrm{KK}^{(i)}\big|_{\infty} \sim c_{i} \, R^{b_{i}} \ ,
\end{equation}
with
\begin{equation}
\label{additional_scaling_parameters}
b_{1} + b_{3} + b_{5} = a_{4} + a_{5} + a_{6} 
\hspace{10mm} \textrm{ and } \hspace{10mm}
\left\lbrace \,\,\,
\begin{array}{ll}   
b_{2} = b_{1} - (1+a_{4}+a_{6}) \\[2mm]
b_{4} = b_{3} - (1+a_{4}+a_{5}) \\[2mm]
b_{6} = b_{5} - (1+a_{5}+a_{6})
\end{array}
\right. \ ,
\end{equation}
so that one finds the useful relation
\begin{equation}
\label{simple_relation_2}
(b_{1}+b_{3}+b_{5}) + (b_{2}+b_{4}+b_{6}) = - 3 \ . 
\end{equation}
Therefore, the $H$-functions are controlled by five scaling parameters $\,(a_{4},a_{5},a_{6} \,;\, b_{1},b_{3})\,$ and six constants $\,(c_{a},c_{i})\,$ in the near-horizon region. Different choices of the free parameters give rise to different AdS$_{3}$ flux vacua with arbitrary gauge fluxes $(f_{7}\,;\,f_{31},\ldots,f_{37})$ and metric fluxes of the form
\begin{equation}
\label{omega_135_fluxes}
\omega_{1} = - \dfrac{a_{4}\,a_{6}\,b_{1}}{c_{\textrm{KK}} \, f_{34}\,f_{36}}
\hspace{5mm} , \hspace{5mm}
\omega_{3} = - \dfrac{a_{4}\,a_{5}\,b_{3}}{c_{\textrm{KK}} \, f_{34}\,f_{35}}
\hspace{5mm} , \hspace{5mm}
\omega_{5} = - \dfrac{a_{5}\,a_{6}\,b_{5}}{c_{\textrm{KK}} \, f_{35}\,f_{36}} \ ,
\end{equation}
and
\begin{equation}
\label{omega_246_fluxes}
\omega_{2} = - \dfrac{a_{5}\, a_{7}\, b_{2}}{c_{\textrm{KK}} \, f_{35}\,f_{37}}
\hspace{5mm} , \hspace{5mm}
\omega_{4} = - \dfrac{a_{6}\, a_{7}\, b_{4}}{c_{\textrm{KK}} \, f_{36}\,f_{37}}
\hspace{5mm} , \hspace{5mm}
\omega_{6} = - \dfrac{a_{4}\, a_{7}\, b_{6}}{c_{\textrm{KK}} \, f_{34}\,f_{37}} \ ,
\end{equation}
where only the product
\begin{equation}
c_{\textrm{KK}} \equiv c_{1}  \, c_{2} \cdots c_{6} \ ,   
\end{equation}
enters the expression for the metric fluxes in (\ref{omega_135_fluxes}) and (\ref{omega_246_fluxes}). As a result, the six metric fluxes $\,(\omega_{a},\omega_{i})\,$ can be expressed in terms of the six independent horizon parameters $(a_{4},a_{5},a_{6} \,;\, b_{1},b_{3} \,;\,c_{\textrm{KK}})$. We will illustrate this with some examples in the next section.

\subsubsection*{Partial near-horizon limits: the background and interpolating solutions}

Let us now consider the general domain-wall solution in (\ref{general_domain_wall}) and take the $H_{5}^{(4,5,6,7)}$-functions and the KK5-monopole functions $\,H^{(a)}_{\textrm{KK}}\,$ and $\,H^{(i)}_{\textrm{KK}}\,$ in their near-horizon limits (\ref{H5_4567_scalings_horizon}) and (\ref{HKK_scalings_horizon}). Then two things can happen:

\begin{itemize}

\item[$i)$] If we move away from the D$1$- and D$5^{(1,2,3)}$-branes by taking the $\,R \rightarrow 0\,$ limit of the harmonic functions in (\ref{H-functions_color}), then
\begin{equation}
\label{H1&H5_123_scalings_asymptotic}
H_{1}\big|_{0} \sim 1 
\hspace{8mm} , \hspace{8mm}
H_{5}^{(1,2,3)}\big|_{0} \sim 1 \ ,
\end{equation}
and the type IIB background solution of Section~\ref{sec:background_solution} is recovered.\footnote{This follows straightforwardly from (\ref{simple_relation_1}) and (\ref{simple_relation_2}), combined with the coordinate redefinition $\,R \sim \hat{y}^{-1}$.}

\item[$ii)$] If the harmonic functions for the D$1$- and D$5^{(1,2,3)}$-branes are kept general as in (\ref{H-functions_color}), then the ten-dimensional interpolating solution of Section~\ref{sec:interpolating_solution} is recovered.

\end{itemize}

\noindent If the near-horizon limit $\,R \rightarrow \infty\,$ is taken also in the harmonic functions of the D$1$- and D$5^{(1,2,3)}$-branes, then the $\,\textrm{AdS}_{3} \times \mathcal{M}_{7}\,$ flux vacuum is consistently recovered.

\subsubsection*{The asymptotically-flat region}

As discussed above, the asymptotic region is identified with $\,R \rightarrow 0\,$ in the solution \eqref{general_domain_wall}. Taking this limit in all the $H$-functions amounts to move away from \textit{all} the sources (which remember sit at $\,R = \infty\,$) in the transverse coordinate $\,R\,$. Apart from setting the harmonic functions in (\ref{H-functions_color}) to
\begin{equation}
H_{1}\big|_{0} \sim 1 
\hspace{5mm} , \hspace{5mm}
H_{5}^{(1)}\big|_{0} \sim 1
\hspace{5mm} , \hspace{5mm}
H_{5}^{(2)}\big|_{0} \sim 1 
\hspace{5mm} , \hspace{5mm}
H_{5}^{(3)}\big|_{0} \sim 1 \ ,
\end{equation}
we must also specify the behaviour of the $H_{5}^{(4,5,6,7)}$-functions when $\,R \rightarrow 0$. A simple \textit{choice}, analogous to \eqref{H-functions_color}, is to assume a minimal power-law behaviour of the form\footnote{One can consider more general $H_{5}^{(k)}$-functions (with $\,k=4,\ldots,7$), for example, a more general scaling behaviour of the form $\,H_{5}^{(k)}\,\sim\, R^{a_{k}\,-\,d_k}\,$ with $\,d_k > 0$, provided the gauge fluxes in (\ref{F3_interpolating_solution_D1-D5}) become constant in the near-horizon region ($R\rightarrow \infty$) and $\,dF_{(3)}\,$ vanishes in the asymptotic region ($R\rightarrow 0$).}
\begin{equation}
\label{H5_456_scalings_asymptotic}
H_{5}^{(4)}\big|_{0} \sim R^{a_{4}-1} 
\hspace{4mm} , \hspace{4mm}
H_{5}^{(5)}\big|_{0} \sim R^{a_{5}-1} 
\hspace{4mm} , \hspace{4mm}
H_{5}^{(6)}\big|_{0} \sim R^{a_{6}-1} 
\hspace{4mm} , \hspace{4mm}
H_{5}^{(7)}\big|_{0} \sim R^{a_{7}-1}  \ ,
\end{equation}
where the $a$'s are the scaling powers in the near-horizon region appearing in (\ref{H5_4567_scalings_horizon}). Then the $H$-functions of the KK$5$-monopoles are obtained by direct integration of (\ref{omega_interpolating_solution_D1-D5}) with constant metric fluxes. In this manner, one finds that the asymptotic spacetime becomes (locally) Riemann-flat -- the ten-dimensional Riemann tensor vanishes when $\,R \rightarrow 0\,$ -- in the orthonormal frame, so no tidal forces are felt by a free falling observer. This is what we will use as a proxy for asymptotic flatness.

\subsection{Full 10D domain-wall solutions}

Putting everything together, the near-horizon and asymptotic regions described above combine into a full 10D domain-wall solution specified by harmonic $H$-functions of the form
\begin{equation}
\label{H-functions_color_full}
H_{1} = 1 - f_{7} \, R 
\hspace{4mm} , \hspace{4mm}
H_{5}^{(1)} = 1 - f_{31} \, R
\hspace{4mm} , \hspace{4mm}
H_{5}^{(2)} = 1 - f_{32} \, R
\hspace{4mm} , \hspace{4mm}
H_{5}^{(3)} = 1 - f_{33} \, R \ ,
\end{equation}
for the D$1$-branes, and (our choice of) non-harmonic $H$-functions of the form
\begin{equation}
\label{H-functions_flavour_full}
\begin{array}{llll}
H_{5}^{(4)} = \dfrac{c_{1}\,c_{3}\,c_{6}}{|a_{4}|} \, R^{a_{4}-1} \,  \Big(  1 + |f_{34}| \, R \Big)
& \hspace{3mm} , \hspace{3mm} &
H_{5}^{(5)} =  \dfrac{c_{2}\,c_{3}\,c_{5}}{|a_{5}|} \, R^{a_{5}-1} \,  \Big(  1 + |f_{35}| \, R \Big) & , \\[6mm]
H_{5}^{(6)} = \dfrac{c_{1}\,c_{4}\,c_{5}}{|a_{6}|} \,  R^{a_{6}-1} \,  \Big(  1 + |f_{36}| \, R \Big)
& \hspace{3mm} , \hspace{3mm} &
H_{5}^{(7)} = \dfrac{c_{2}\,c_{4}\,c_{6}}{|a_{7}|} \, R^{a_{7}-1} \,  \Big(  1 + |f_{37}| \, R \Big) & ,
\end{array}
\end{equation}
for the D$5$-branes in the D1-D5-KK5 intersection of Table~\ref{tab:DW-branes}. The $H$-functions encoding the KK$5$-monopoles are then obtained by direct integration of (\ref{omega_interpolating_solution_D1-D5}) with \textit{constant} metric fluxes $\,(\omega_{a},\omega_{i})$. We will present the complete ten-dimensional domain-wall solution for various representative examples of scale-separated AdS$_{3}$ flux vacua collected in Appendix~\ref{app:AdS3_vacua}.

\subsection{D1-D5-KK5$^{(1)}$ intersection}
\label{sec:example_KK1}

A first class of supersymmetric AdS$_{3}$ flux vacua arises when the metric fluxes originate solely from the KK5$^{(1,a)}$ monopoles listed in Table~\ref{tab:DW-branes}. The presence of only this class of KK$5$-monopoles allows the configurations to be mapped, after three T-dualities along the $\,\eta^{a}\,$ directions, to massless type IIA backgrounds including NS5$^{(1,a)}$-branes, provided the fluxes transform according to the mapping
\begin{equation}
\label{IIA_dual_KK1}
f_{ijkabc7} \rightarrow f_{ijk7}
\hspace{3mm} , \hspace{3mm}
f_{ai7} \rightarrow f_{ibc7}
\hspace{3mm} , \hspace{3mm}
f_{abk} \rightarrow f_{ck}
\hspace{3mm} , \hspace{3mm}
f_{ijk} \rightarrow f_{abcijk}
\hspace{3mm} , \hspace{3mm}
\omega_{i7}{}^{a} \rightarrow H_{ai7} \ .
\end{equation}
For the sake of simplicity, we will set
\begin{equation}
\label{metric_fluxes_example_KK1}
\omega_{1} = \omega_{3} = \omega_{5} \equiv \omega
\hspace{10mm} , \hspace{10mm}
\omega_{2} = \omega_{4} = \omega_{6} = 0 \ ,
\end{equation}
so we will include the same number of  $\,\textrm{KK}5^{(1,a)}\equiv\textrm{KK}5^{(1)}\,$ monopoles in the compactification scheme. In this case, the basis of one-forms in (\ref{one-form_integrated_general}) simplifies to
\begin{equation}
\label{one-form_integrated_KK1}
\eta^{a} = dx^{a} + \omega \, x^{7} \, dx^{i}
\hspace{8mm} , \hspace{8mm}
\eta^{i}  = dx^{i}
\hspace{8mm} \textrm{ and } \hspace{8mm}
\eta^{7}  = dx^{7} \ ,
\end{equation}
and the group manifold becomes a $7$-dimensional two-step nilmanifold. The isometry algebra of the nilmanifold consists of the direct sum of three copies of the 3D Heisenberg algebra sharing the same 7$^{\textrm{th}}$ generator (base). The flux parameters in $\,F_{(3)}\,$ entering the tadpole cancellation conditions in (\ref{Tadpole_O5/D5_4567}) are taken as
\begin{equation}
\label{gauge_fluxes_example_KK1}
f_{34} = f_{35} = f_{36} \equiv \tilde{f} > 0 
\hspace{8mm} \textrm{ and } \hspace{8mm}
f_{37} < 0  \ ,
\end{equation}
so three out of the four tadpoles in (\ref{Tadpole_O5/D5_4567}) are non-zero and contribute to the scalar potential as
\begin{equation}
\label{V_sources_example_KK1}
V_{\widetilde{\textrm{O}5}/\widetilde{\textrm{D}5}^{(4)}}=V_{\widetilde{\textrm{O}5}/\widetilde{\textrm{D}5}^{(5)}} =  V_{\widetilde{\textrm{O}5}/\widetilde{\textrm{D}5}^{(6)}}  = - 2 \, \omega \, \tilde{f} < 0
\hspace{5mm} \textrm{ and } \hspace{5mm}
V_{\widetilde{\textrm{O}5}/\widetilde{\textrm{D}5}^{(7)}} = 0 \ .
\end{equation}
Therefore, $\,\widetilde{\textrm{O}5}^{(4,5,6)}$-planes dominate over $\widetilde{\textrm{D}5}^{(4,5,6)}$-branes in agreement with \cite{Tringas:2025uyg}. Relevant data about this family of AdS$_{3}$ flux vacua is collected in Appendix~\ref{sec:AdS3_KK_1}.

\subsubsection*{The $H$-functions}

The metric and gauge fluxes in (\ref{metric_fluxes_example_KK1}) and (\ref{gauge_fluxes_example_KK1}) completely fix the scaling parameters to take the values
\begin{equation}
a_{4} = -1
\hspace{4mm} , \hspace{4mm}
a_{5} = -1
\hspace{4mm} , \hspace{4mm}
a_{6} = -1
\hspace{4mm} , \hspace{4mm}
b_{1}=-1
\hspace{4mm} , \hspace{4mm}
b_{3}=-1 \ .
\end{equation}
Using the relations in (\ref{additional_scaling_parameters}) and demanding a constant metric flux $\,\omega\,$ from (\ref{omega_interpolating_solution_D1-D5}), one arrives at the following 10D domain-wall solution with the correct behaviour at $\,R \rightarrow \infty\,$ and $\,R \rightarrow 0$. The $H$-functions for the D$1$-branes and the D$5$-branes in (\ref{H-functions_color_full}) and (\ref{H-functions_flavour_full}) are given by
\begin{equation}
\label{H-functions_color_full_KK1}
H_{1} = 1 - f_{7} \, R 
\hspace{4mm} , \hspace{4mm}
H_{5}^{(1)} = 1 - f_{31} \, R
\hspace{4mm} , \hspace{4mm}
H_{5}^{(2)} = 1 - f_{32} \, R
\hspace{4mm} , \hspace{4mm}
H_{5}^{(3)} = 1 - f_{33} \, R \ ,
\end{equation}
and
\begin{equation}
\label{H-functions_flavour_full_KK1}
\begin{array}{llll}
H_{5}^{(4)} = c_{1}\,c_{3}\,c_{6} \, R^{-2}  \,  \Big(  1 + \tilde{f}\,R \Big)
& \hspace{3mm} , \hspace{3mm} &
H_{5}^{(5)} =  c_{2}\,c_{3}\,c_{5} \, R^{-2}  \,  \Big(  1 + \tilde{f}\,R \Big) & , \\[6mm]
H_{5}^{(6)} =  c_{1}\,c_{4}\,c_{5} \, R^{-2}  \,  \Big(  1 + \tilde{f}\,R \Big)
& \hspace{3mm} , \hspace{3mm} &
H_{5}^{(7)} =  c_{2}\,c_{4}\,c_{6}  \,  \Big(  1 - f_{37}\,R \Big) & .
\end{array}
\end{equation}
Note that the functions $\,H_{5}^{(4,5,6)}\,$ turn out \textit{not} to be harmonic and the associated domain-wall branes are smeared. There are non-zero flux-induced tadpoles in (\ref{Tadpole_O5/D5_4567}) for the $\widetilde{\textrm{O}5}$/$\widetilde{\textrm{D}5}^{(4,5,6)}$ brane sources at the horizon, which therefore contribute as (\ref{V_sources_example_KK1}) to the scalar potential. The $H$-functions encoding the KK$5$-monopoles take the form
\begin{equation}
H_\textrm{KK}^{(2)} = c_{2} 
\hspace{6mm} , \hspace{6mm}
H_\textrm{KK}^{(4)} = c_{4} 
\hspace{6mm} , \hspace{6mm}
H_\textrm{KK}^{(6)} = c_{6} \ ,
\end{equation}
together with
\begin{equation}
\label{HKK_scalings_full_KK1}
H_\textrm{KK}^{(1)} = - \, \omega \, c_{2} \int  H_{5}^{(4)}  H_{5}^{(6)} 
\hspace{3mm} , \hspace{3mm}
H_\textrm{KK}^{(3)} = - \, \omega \, c_{4} \int  H_{5}^{(4)}  H_{5}^{(5)} 
\hspace{3mm} , \hspace{3mm}
H_\textrm{KK}^{(5)} = - \, \omega \, c_{6} \int  H_{5}^{(5)}  H_{5}^{(6)} \ .
\end{equation}
Finally, in order to recover the AdS$_{3}$ flux vacuum in the near-horizon region $\,R\rightarrow \infty$, the arbitrary constants $\,(c_{a},c_{i})\,$ must satisfy the single relation
\begin{equation}
c_{\textrm{KK}} =   \dfrac{1}{\omega \, \tilde{f}^{2}} \ .
\end{equation}
The Riemann tensor for a free-falling observer falls off as $\,R_{ABCD}\sim R^{10}\,$ in the asymptotic region $\,R\rightarrow 0$, so the ten-dimensional space-time becomes (locally) asymptotically flat.

\subsection{D1-D5-KK5$^{(2)}$ intersection}
\label{sec:example_KK2}

A second class of AdS$_{3}$ flux vacua can be realised in the presence of metric fluxes sourced exclusively by the KK5$^{(2,i)}$ monopoles listed in Table~\ref{tab:DW-branes}. These configurations are T-dual -- after three T-dualities along the $\,\eta^{i}\,$ directions -- to massive type IIA backgrounds including NS5$^{(2,i)}$-branes provided the flux mapping\footnote{They belong to the class of massive IIA compactifications on G$_2$-orientifolds investigated in \cite{Farakos:2020phe}.}
\begin{equation}
\label{IIA_dual_KK2}
f_{ijkabc7} \rightarrow f_{abc7}
\hspace{3mm} , \hspace{3mm}
f_{ai7} \rightarrow f_{ajk7}
\hspace{3mm} , \hspace{3mm}
f_{abk} \rightarrow f_{aibj}
\hspace{3mm} , \hspace{3mm}
f_{ijk} \rightarrow f_{0}
\hspace{3mm} , \hspace{3mm}
\omega_{a7}{}^{i} \rightarrow H_{ia7} \ .
\end{equation}
Let us make again the simple choice
\begin{equation}
\label{metric_fluxes_example_KK2}
\omega_{1} = \omega_{3} = \omega_{5} = 0 
\hspace{10mm} , \hspace{10mm}
\omega_{2} = \omega_{4} = \omega_{6} \equiv \omega \ ,
\end{equation}
so that the same number of  $\,\textrm{KK}5^{(2,i)}\equiv \textrm{KK}5^{(2)}\,$ monopoles is included. The basis of one-forms in (\ref{one-form_integrated_general}) simplifies to
\begin{equation}
\label{one-form_integrated_KK2}
\eta^{i} = dx^{i} - \omega \, x^{7} \, dx^{a}
\hspace{8mm} , \hspace{8mm}
\eta^{a}  = dx^{a}
\hspace{8mm} \textrm{ and } \hspace{8mm}
\eta^{7}  = dx^{7} \ ,
\end{equation}
and, as in the previous example, the group manifold becomes a nilmanifold with an isometry algebra described again by the direct sum of three copies of the 3D Heisenberg algebra sharing the same 7$^{\textrm{th}}$ generator. The flux parameters in $\,F_{(3)}\,$ entering the tadpole cancellation conditions are now taken to be
\begin{equation}
\label{gauge_fluxes_example_KK2}
f_{34} = f_{35} = f_{36} = 0 
\hspace{8mm} \textrm{ and } \hspace{8mm}
f_{37} > 0  \ ,
\end{equation}
so three out of the four tadpoles in (\ref{Tadpole_O5/D5_4567}) are non-zero. These induce contributions to the scalar potential of the form
\begin{equation}
\label{V_sources_example_KK2}
V_{\widetilde{\textrm{O}5}/\widetilde{\textrm{D}5}^{(4)}}=V_{\widetilde{\textrm{O}5}/\widetilde{\textrm{D}5}^{(5)}} = V_{\widetilde{\textrm{O}5}/\widetilde{\textrm{D}5}^{(6)}}= -  \omega \, f_{37} < 0    
\hspace{5mm} \textrm{ and } \hspace{5mm}
V_{\widetilde{\textrm{O}5}/\widetilde{\textrm{D}5}^{(7)}} = 0 \ ,
\end{equation}
so, as before, $\,\widetilde{\textrm{O}5}^{(4,5,6)}$-planes dominate over $\widetilde{\textrm{D}5}^{(4,5,6)}$-branes. We have collected the relevant data about this family of AdS$_{3}$ flux vacua in Appendix~\ref{sec:AdS3_KK_2}.

One distinctive feature of this specific example of AdS$_{3}$ flux vacua is that not all the moduli are stabilised. Nonetheless, full moduli stabilisation can be achieved by turning on generic $\,(f_{34},f_{35},f_{36})\,$ fluxes in (\ref{gauge_fluxes_example_KK2}). In that case, the scenario becomes essentially identical to the previous example, so we will focus here on the case with unstabilised moduli.

\subsubsection*{The $H$-functions}

The metric and gauge fluxes in (\ref{metric_fluxes_example_KK2}) and (\ref{gauge_fluxes_example_KK2}) fix the scaling parameters to be
\begin{equation}
a_{4} = 0
\hspace{4mm} , \hspace{4mm}
a_{5} = 0
\hspace{4mm} , \hspace{4mm}
a_{6} = 0
\hspace{4mm} , \hspace{4mm}
b_{1}=0
\hspace{4mm} , \hspace{4mm}
b_{3}=0 \ .
\end{equation}
Using the relations in (\ref{additional_scaling_parameters}), the $H$-functions for the D$1$-branes and D$5$-branes in (\ref{H-functions_color_full}) and (\ref{H-functions_flavour_full}) are given by
\begin{equation}
\label{H-functions_color_full_KK2}
H_{1} = 1 - f_{7} \, R 
\hspace{4mm} , \hspace{4mm}
H_{5}^{(1)} = 1 - f_{31} \, R
\hspace{4mm} , \hspace{4mm}
H_{5}^{(2)} = 1 - f_{32} \, R
\hspace{4mm} , \hspace{4mm}
H_{5}^{(3)} = 1 - f_{33} \, R \ ,
\end{equation}
and
\begin{equation}
\label{H-functions_flavour_full_KK2}
\begin{array}{llll}
H_{5}^{(4)} = \dfrac{c_{1}\,c_{3}\,c_{6}}{2}
& \hspace{3mm} , \hspace{3mm} &
H_{5}^{(5)} =  \dfrac{c_{2}\,c_{3}\,c_{5}}{2}  , \\[6mm]
H_{5}^{(6)} = \dfrac{c_{1}\,c_{4}\,c_{5}}{2} 
& \hspace{3mm} , \hspace{3mm} &
H_{5}^{(7)} =  \dfrac{c_{2}\,c_{4}\,c_{6}}{2} \, R^{-3} \,  \Big(  1 + f_{37} \,R \Big) & ,
\end{array}
\end{equation}
so, in contrast to the previous example, only the $\,H_{5}^{(7)}$-function happens \textit{not} to be harmonic. There are again three non-zero flux-induced tadpoles associated with the $\widetilde{\textrm{O}5}$/$\widetilde{\textrm{D}5}^{(4,5,6)}$ brane sources at the horizon, which produce the three contributions to the scalar potential in (\ref{V_sources_example_KK2}). Lastly, the $H$-functions encoding the KK$5$-monopoles are obtained from (\ref{omega_interpolating_solution_D1-D5}), namely,
\begin{equation}
H_\textrm{KK}^{(1)} = c_{1} 
\hspace{6mm} , \hspace{6mm}
H_\textrm{KK}^{(3)} = c_{3} 
\hspace{6mm} , \hspace{6mm}
H_\textrm{KK}^{(5)} = c_{5} \ ,
\end{equation}
together with
\begin{equation}
\label{HKK_scalings_full_KK2}
H_\textrm{KK}^{(2)} = \omega \, c_{1} \int  H_{5}^{(5)}  H_{5}^{(7)} 
\hspace{3mm} , \hspace{3mm}
H_\textrm{KK}^{(4)} = \omega \, c_{3} \int  H_{5}^{(6)}  H_{5}^{(7)} 
\hspace{3mm} , \hspace{3mm}
H_\textrm{KK}^{(6)} = \omega \, c_{5} \int  H_{5}^{(4)}  H_{5}^{(7)} \ .
\end{equation}
Recovering the AdS$_{3}$ flux vacuum in the near-horizon region $\,R\rightarrow \infty\,$ imposes again a single relation of the form
\begin{equation}
c_{\textrm{KK}}  =  \dfrac{4 \, \kappa}{\omega \, \sqrt{f_{37}} } \ ,
\end{equation}
and the Riemann tensor for a free-falling observer goes to zero as $\,R_{ABCD}\sim R^{\frac{11}{2}}\,$ in the asymptotic region $\,R\rightarrow 0$, thus recovering an asymptotically-flat spacetime (locally).

\subsection{D1-D5-KK5$^{(1)}$-KK5$^{(2)}$ intersection}
\label{sec:example_KK1&KK2}

A third class of supersymmetric AdS$_{3}$ flux vacua can be considered which simultaneously requires the two types of KK$5$-monopoles in Table~\ref{tab:DW-branes}. This is a variant of the setup in \cite{Arboleya:2024vnp,VanHemelryck:2025qok} with metric fluxes of the form
\begin{equation}
\label{metric_fluxes_example_KK1&KK2}
\omega_{1} = \omega_{3} = \omega_{5} \equiv \omega
\hspace{10mm} , \hspace{10mm}
\omega_{2} = \omega_{4} = \omega_{6} \equiv \omega \ ,
\end{equation}
so the group manifold is the $7$-dimensional solvmanifold with one-form basis elements (\ref{one-form_integrated_general}). Unlike the previous examples, this flux configuration cannot be T-dualised into a geometric type IIA compactification. For illustrative purposes, let us make a simple choice of the flux parameters in $\,F_{(3)}\,$ that enter the tadpole cancellation conditions in (\ref{Tadpole_O5/D5_4567}). We take them equal, namely,
\begin{equation}
\label{gauge_fluxes_example_KK1&KK2}
f_{34} = f_{35} = f_{36} = f_{37} \equiv \tilde{f} > 0 \ ,
\end{equation}
so all the flux-induced tadpoles in (\ref{Tadpole_O5/D5_4567}) are equal and non-zero. In particular, they contribute to the scalar potential as
\begin{equation}
\label{V_sources_example_KK1_KK2}
V_{\widetilde{\textrm{O}5}/\widetilde{\textrm{D}5}^{(4)}}=V_{\widetilde{\textrm{O}5}/\widetilde{\textrm{D}5}^{(5)}} = V_{\widetilde{\textrm{O}5}/\widetilde{\textrm{D}5}^{(6)}} = V_{\widetilde{\textrm{O}5}/\widetilde{\textrm{D}5}^{(7)}} = - 3 \, \omega \, \tilde{f} < 0 \ ,
\end{equation}
so $\widetilde{\textrm{O}5}$-planes dominate over $\widetilde{\textrm{D}5}$-branes in agreement with \cite{Tringas:2025uyg}. Relevant data about this family of AdS$_{3}$ flux vacua is collected in Appendix~\ref{sec:AdS3_KK_1&KK_2}.

\subsubsection*{The $H$-functions}

The scaling parameters required to reproduce the metric and gauge fluxes in (\ref{metric_fluxes_example_KK1&KK2}) and (\ref{gauge_fluxes_example_KK1&KK2}) are given by
\begin{equation}
a_{4} = -\frac{1}{2}
\hspace{4mm} , \hspace{4mm}
a_{5} = -\frac{1}{2}
\hspace{4mm} , \hspace{4mm}
a_{6} = -\frac{1}{2}
\hspace{4mm} , \hspace{4mm}
b_{1}= -\frac{1}{2}
\hspace{4mm} , \hspace{4mm}
b_{3}= -\frac{1}{2} \ .
\end{equation}
Using the relations (\ref{additional_scaling_parameters}), the $H$-functions for the D$1$-branes and D$5$-branes in (\ref{H-functions_color_full}) and (\ref{H-functions_flavour_full}) that recover $\,\textrm{AdS}_{3}\times\mathcal{M}_{7}\,$ at the horizon ($R\rightarrow\infty$) and asymptotic Riemann flatness at $\,R\rightarrow 0\,$ take the form
\begin{equation}
\label{H-functions_color_full_KK1&KK2}
H_{1} = 1 - f_{7} \, R 
\hspace{4mm} , \hspace{4mm}
H_{5}^{(1)} = 1 - f_{31} \, R
\hspace{4mm} , \hspace{4mm}
H_{5}^{(2)} = 1 - f_{32} \, R
\hspace{4mm} , \hspace{4mm}
H_{5}^{(3)} = 1 - f_{33} \, R \ ,
\end{equation}
and
\begin{equation}
\label{H-functions_flavour_full_KK1&KK2}
\begin{array}{llll}
H_{5}^{(4)} = 2\,c_{1}\,c_{3}\,c_{6} \, R^{-3/2} \,  \Big(  1 + \tilde{f}\,R \Big)
& \hspace{3mm} , \hspace{3mm} &
H_{5}^{(5)} =  2\,c_{2}\,c_{3}\,c_{5} \, R^{-3/2} \,  \Big(  1 +  \tilde{f}\,R \Big) & , \\[6mm]
H_{5}^{(6)} = 2\,c_{1}\,c_{4}\,c_{5} \,  R^{-3/2} \,  \Big(  1 + \tilde{f}\, R \Big)
& \hspace{3mm} , \hspace{3mm} &
H_{5}^{(7)} = 2\,c_{2}\,c_{4}\,c_{6}\, R^{-3/2} \,  \Big(  1 + \tilde{f} \,R \Big) & .
\end{array}
\end{equation}
Now all the functions $\,H_{5}^{(4,5,6,7)}\,$ turn out \textit{not} to be harmonic. The four flux-induced tadpoles in (\ref{Tadpole_O5/D5_4567}) for the $\widetilde{\textrm{O}5}$/$\widetilde{\textrm{D}5}^{(4,5,6,7)}$ brane sources at the horizon are non-zero, which contribute to the scalar potential as in (\ref{V_sources_example_KK1_KK2}). Demanding the metric flux $\,\omega\,$ in (\ref{omega_interpolating_solution_D1-D5}) to be constant fixes the $H$-functions of the KK$5$-monopoles to be
\begin{equation}
\label{HKK_scalings_full_KK1&KK2}
H_\textrm{KK}^{(a)} = c_{a}\,R^{-1/2}\,e^{\frac{1+4\,\tilde{f}\,R}{4\,\tilde{f}^2\,R^2}} \hspace{6mm} , \hspace{6mm}  H_\textrm{KK}^{(i)} = c_{i}\,R^{-1/2}\,e^{\frac{1+4\,\tilde{f}\,R}{4\,\tilde{f}^2\,R^2}}\,.
\end{equation}
Recovering the AdS$_3$ vacuum in the near-horizon region $\,R\rightarrow\infty\,$ then only imposes
\begin{equation}
c_{\textrm{KK}} =  \dfrac{1}{8\,\omega \, \tilde{f}^2} \ ,
\end{equation}
and the Riemann tensor in the orthonormal frame behaves as $\,R_{ABCD}\sim e^{-1/R^{2}}\,$ for $\,R \rightarrow 0$, thus indicating that the ten-dimensional spacetime becomes (locally) asymptotically flat.

\subsection{On the flux/brane correspondence}

According to the flux/brane correspondence \cite{Kounnas:2007dd}, in the near-horizon region the domain-wall branes are replaced by the constant fluxes that support the AdS vacuum, while the presence of the source branes is encoded in the tadpole cancellation conditions satisfied by the resulting flux background. Here we discuss how this correspondence is realised in the class of scale-separated AdS$_3$ flux vacua investigated in this work.

\begin{figure}[t]
\begin{center}
\includegraphics[width=0.8\textwidth]{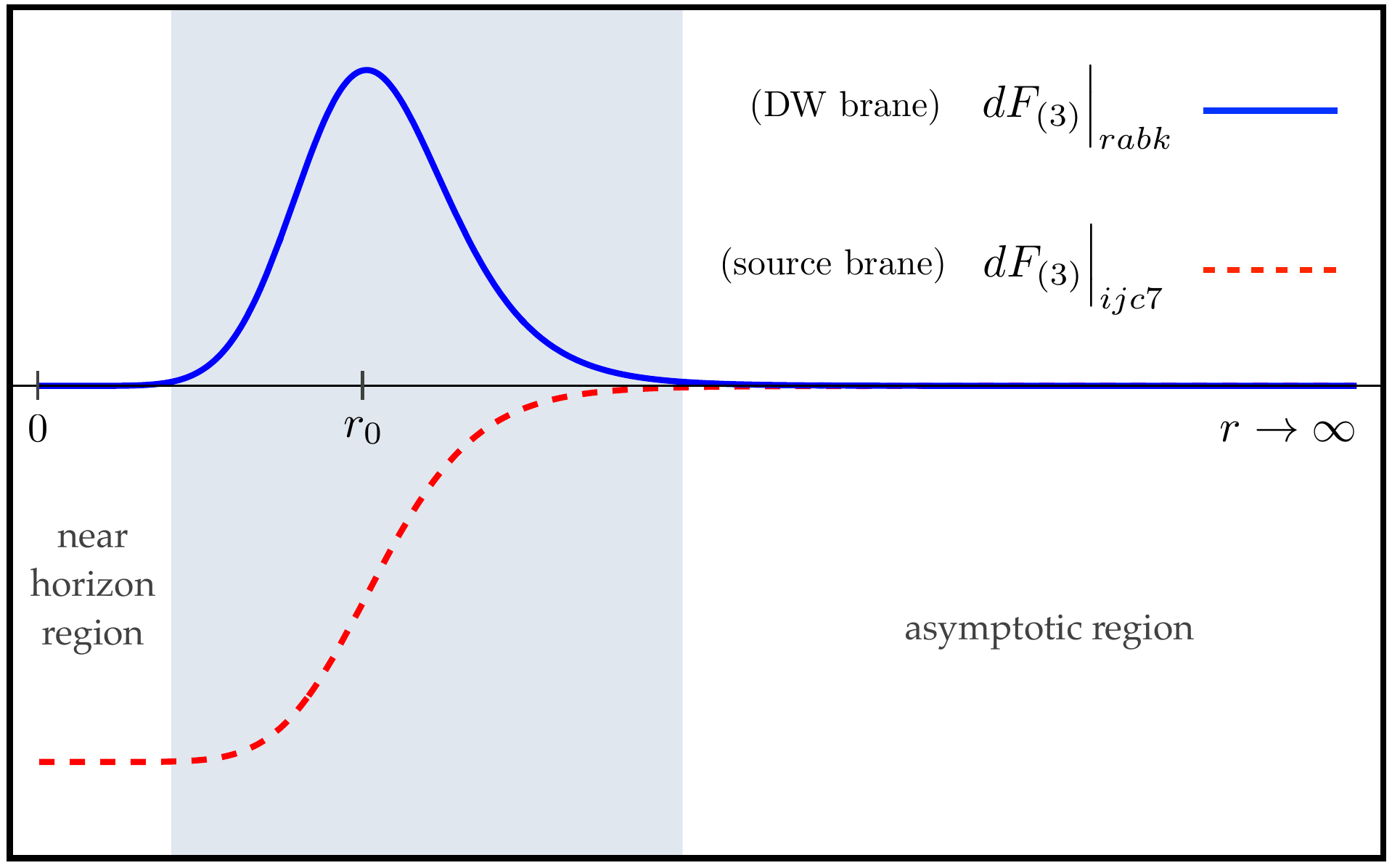} 
\end{center}
\caption{Flux/brane correspondence for the example in Section~\ref{sec:example_KK1}. In the near-horizon region ($r \rightarrow 0$) only source branes are present in agreement with recovering an $\,\textrm{AdS}_{3} \times \mathcal{M}_{7}\,$ flux vacuum with non-zero flux-induced tadpoles. In the asymptotic region ($r \rightarrow \infty$) there are neither domain-wall (DW) nor source branes in agreement with recovering flat space-time. The transition between the two regions occurs in the shaded zone around $\,r_{0}\,$ where both types of branes are present and smeared as dictated by our choice of $H$-functions in (\ref{H-functions_flavour_full_KK1}). This figure has been generated with all parameters ($\omega$, $\tilde{f}$, ...) set to unity.}
\label{Fig:brane-flux_transition_example_KK1}
\end{figure}

For concreteness, we focus the remainder of the discussion on the example presented in Section~\ref{sec:example_KK1}. The $H$-functions were chosen as in (\ref{H-functions_color_full_KK1})-(\ref{HKK_scalings_full_KK1}) involving non-harmonic functions $\,H_{5}^{(4,5,6)}\,$ for the $\textrm{O}5$/$\textrm{D}5^{(4,5,6)}$ domain-wall branes in Table~\ref{tab:DW-branes}. Non-zero tadpoles were induced for the $\widetilde{\textrm{O}5}$/$\widetilde{\textrm{D}5}^{(4,5,6)}$ spacetime-filling brane sources in Table~\ref{tab:sources}, which support the AdS$_{3}$ flux vacuum arising at the near-horizon region. These two types of branes enter the ten-dimensional Bianchi identities as
\begin{equation}
dF_{(3)} \Big|_{rabk} = J^{\textrm{O}5/\textrm{D}5}_{(4,5,6)}
\hspace{12mm} \textrm{ and } \hspace{12mm}
dF_{(3)} \Big|_{ijc7} = \tilde{J}^{\widetilde{\textrm{O}5}/\widetilde{\textrm{D}5}}_{(4,5,6)} \ ,
\end{equation}
where (for later convenience in Figure~\ref{Fig:brane-flux_transition_example_KK1}) we have introduced a new transverse coordinate $\,r = R^{-1/4}$, and where the $J$'s and $\tilde{J}$'s are the smearing functions in (\ref{currents_branes}). By tracking the evolution of $\,dF_{(3)} \Big|_{rabk} \sim \partial_{[r} F_{abk]}\,$ and $\,dF_{(3)} \Big|_{ijc7} \sim \omega_{[i7}{}^{a} F_{jc]a}\,$ along the transverse coordinate, we can monitor the emergence of the flux vacuum in the near-horizon region.\footnote{Recall that, for the unrestricted D1- and D5$^{(1,2,3)}$-branes, one has $\,dF_{(7)} = 0\,$ as well as $\,\left. dF_{(3)} \right|_{rai7}=0\,$ and $\,\left. dF_{(3)} \right|_{bjck}=0$ all along the transverse coordinate $\,r$. For the example of Section~\ref{sec:example_KK1}, the same happens for the D5$^{(7)}$-branes, namely, $\,\left. dF_{(3)} \right|_{r246}=0\,$ and $\,\left. dF_{(3)} \right|_{1357}=0\,$ all along the transverse coordinate $\,r$.} This is depicted in Figure~\ref{Fig:brane-flux_transition_example_KK1} for the example in Section~\ref{sec:example_KK1}. Nonetheless, let us emphasise again that the non-harmonic $H$-functions in (\ref{H-functions_flavour_full}) -- or in (\ref{H-functions_flavour_full_KK1}) for the particular example in Section~\ref{sec:example_KK1} -- represent just one possible choice. Different choices give rise to different brane smearings in the shaded zone of Figure~\ref{Fig:brane-flux_transition_example_KK1} where the transition to the flux vacuum occurs.

Finally, as established in Section~\ref{sec:example_KK1}, there is an alternative T-dual description of this brane system in massless type IIA supergravity in which fluxes are mapped as in (\ref{IIA_dual_KK1}), and the $\textrm{O}5$/$\textrm{D}5^{(4,5,6)}$ domain-wall branes and $\widetilde{\textrm{O}5}$/$\widetilde{\textrm{D}5}^{(4,5,6)}$ brane sources are mapped into the $\textrm{O}6$/$\textrm{D}6^{(4,5,6)}$ domain-wall branes and $\widetilde{\textrm{O}4}$/$\widetilde{\textrm{D}4}^{(4,5,6)}$ brane sources summarised in Table~\ref{tab:IIA_branes}.

\begin{table}[t]
\centering
\renewcommand{\arraystretch}{1.5}
\begin{tabular}{|c||cc||c||cc|cc|cc||c|}
\hline
& $dt$ & $dx$ & $dy$ & $\eta^{1}$  & $\eta^{2}$ & $\eta^{3}$  & $\eta^{4}$ & $\eta^{5}$ & $\eta^{6}$ & $\eta^{7}$  \\
\hline
\hline
$\textrm{O}6$/$\textrm{D}6^{(4)}$ & $\times$ & $\times$ &  & $\times$ & $\times$ & $\times$ & $\times$ &  &   & $\times$  \\
$\textrm{O}6$/$\textrm{D}6^{(5)}$ & $\times$ & $\times$ &  &  &   & $\times$ & $\times$ & $\times$  & $\times$ & $\times$\\ 
$\textrm{O}6$/$\textrm{D}6^{(6)}$ & $\times$ & $\times$ &  & $\times$  & $\times$ & &   &  $\times$ & $\times$ &  $\times$  \\
\hline
\hline
$\widetilde{\textrm{O}4}$/$\widetilde{\textrm{D}4}^{(4)}$ & $\times$ & $\times$ & $\times$ &  &  &  &  & $\times$  & $\times$  &  \\
$\widetilde{\textrm{O}4}$/$\widetilde{\textrm{D}4}^{(5)}$ & $\times$ & $\times$ & $\times$ & $\times$  & $\times$  &  &  &  &  &  \\ 
$\widetilde{\textrm{O}4}$/$\widetilde{\textrm{D}4}^{(6)}$ & $\times$ & $\times$ & $\times$ &  &  & $\times$ & $\times$  &  &  &   \\
\hline
\end{tabular}
\caption{Domain-wall branes (upper) and spacetime-filling sources (lower) entering ten-dimensional Bianchi identities $\,dF_{(2)} \Big|_{rai} = J^{\textrm{O}6/\textrm{D}6}_{(4,5,6)}\,$ and $\,
- H_{3} \wedge F_{2} \Big|_{bjck7} = \tilde{J}^{\widetilde{\textrm{O}4}/\widetilde{\textrm{D}4}}_{(4,5,6)}\,$ in the T-dual (massless) type IIA description of the AdS$_{3}$ flux vacua in Section~\ref{sec:example_KK1}.}
\label{tab:IIA_branes}
\end{table}

\section{Summary and final remarks}
\label{sec:discussion}

We have investigated the brane origin of the supersymmetric and scale-separated AdS$_3$ flux vacua that arise from a class of type IIB co-calibrated G$_2$-orientifolds (see \cite{Emelin:2021gzx}\cite{Arboleya:2024vnp,VanHemelryck:2025qok,Arboleya:2025ocb}). After some partial results obtained using the flux backtracking method of \cite{Apers:2025pon,Apers:2026lgi}, we constructed the D1-D5-KK5 intersection whose near-horizon region realises the general scale-separated AdS$_3$ flux vacua in (\ref{general_AdS3_1})-(\ref{general_AdS3_2}), thus establishing a connection between these three-dimensional flux vacua and their higher-dimensional smeared brane origin.

For illustrative purposes, we presented three examples, each representing a different scenario. In all of them, the D1-D5-KK5 intersection approaches a (locally) Riemann-flat ten-dimensional spacetime in the asymptotic region.

\paragraph{1) Flux vacua with a massless type IIA dual.} In Section~\ref{sec:example_KK1} we considered supersymmetric AdS$_3$ flux vacua supported by a single class of KK$5$-monopoles, namely the KK5$^{(1,a)}$-monopoles of Table~\ref{tab:DW-branes}, and hence by only one type of metric flux, $\,\omega_a > 0$. The corresponding seven-dimensional internal space is constructed from the nilmanifold (\ref{one-form_integrated_KK1}) (after compact quotient)\footnote{In \cite{Andriot:2016rdd,Andriot:2018tmb}, the complete spectrum (eigenforms and eigenvalues) of the Laplacian on the
three-dimensional Heisenberg nilmanifold was determined. It would be interesting to carry out a similar spectral analysis for the seven-dimensional nilmanifold (\ref{one-form_integrated_KK1}) in order to better understand whether a Scherk--Schwarz Ansatz based on this nilmanifold determines not only a consistent truncation but also an effective field theory.}. This type IIB scenario is T-dual to a massless type IIA scenario including NS5$^{(1,a)}$-branes through the flux mapping (\ref{IIA_dual_KK1}), and suffices to stabilise the eight dilatons in the G$_{2}$-compactification. For the simple choice of fluxes in (\ref{gauge_fluxes_example_KK1}), the brane intersection in (\ref{general_domain_wall})-(\ref{F3_interpolating_solution_D1-D5}) is determined by the $H$-functions in (\ref{H-functions_color_full_KK1})-(\ref{HKK_scalings_full_KK1}). The $H_{5}^{(4,5,6)}$-functions are non-harmonic, the associated domain-wall branes are smeared (see Figure~\ref{Fig:brane-flux_transition_example_KK1}), and there are non-zero flux-induced tadpoles for the $\widetilde{\textrm{O}5}/\widetilde{\textrm{D}5}^{(4,5,6)}$ brane sources (see Table~\ref{tab:sources}) at the AdS$_{3}$ flux vacuum that is recovered in the near-horizon region. This AdS$_{3}$ vacuum comes along with a low-lying spectrum of dual operators of the form
\begin{equation}
\Delta =
6 \,\, (\times 1) 
\,\,\,,\,\,\, 
4 \,\, (\times 7) \ ,
\end{equation}
so no extremal ($n=0$) or over-extremal ($n \neq 0$) cubic couplings between operators satisfying $\,\Delta_{i} + \Delta_{j} = \Delta_{k} - 2 \, n\,$ with $\,n \in \mathbb{N}\,$ are possible, thus passing the consistency constraint put forward in \cite{Bobev:2025yxp}.

\paragraph{2) Flux vacua with a massive type IIA dual.} In Section~\ref{sec:example_KK2} we considered supersymmetric AdS$_3$ flux vacua supported this time only by the KK5$^{(2,i)}$-monopoles of Table~\ref{tab:DW-branes}. Namely, supported only by the type of metric flux $\,\omega_i > 0$. The internal manifold is constructed (after compact quotient) from the nilmanifold (\ref{one-form_integrated_KK2}). Upon three T-dualities, this type IIB scenario becomes a massive IIA scenario including NS5$^{(2,i)}$-branes provided the flux mapping in (\ref{IIA_dual_KK2}). As in the previous example, generic flux choices stabilise all eight dilatons of the G$_2$-compactification in an AdS$_3$ flux vacuum. We nevertheless focused on the special flux configuration in (\ref{gauge_fluxes_example_KK2}), for which some moduli remain unfixed. The resulting AdS$_3$ solutions therefore form continuous loci in moduli space rather than isolated vacua. Even in this case, both the flux-backtracking procedure \cite{Apers:2025pon,Apers:2026lgi} and the D1-D5-KK5 intersection we propose recover such continuous class of AdS$_{3}$ vacua in the near-horizon region. Only the $H_{5}^{(7)}$-function in (\ref{H-functions_flavour_full_KK2}) is non-harmonic, and a non-zero tadpole is induced for the $\widetilde{\textrm{O}5}/\widetilde{\textrm{D}5}^{(4,5,6)}$ brane sources (see Table~\ref{tab:sources}) at the AdS$_{3}$ flux vacuum. The dimensions of the low-lying operators in the putative dual CFT now read
\begin{equation}
\Delta = 
8 \,\, (\times 1) 
\,\,\,,\,\,\, 
4 \,\, (\times 4) 
\,\,\,,\,\,\, 
2 \,\, (\times 3) \ ,
\end{equation}
thus allowing for the potentially dangerous arrangements $\,\Delta_{i} + \Delta_{j} = \Delta_{k} - 2 \, n\,$ with $\,n \in \mathbb{N}\,$ pointed out in \cite{Bobev:2025yxp}. Being a peculiarity of the flux choice we made, we are not exploring this issue any further.

\paragraph{3) Flux vacua without a type IIA dual.} In Section~\ref{sec:example_KK1&KK2} we considered supersymmetric AdS$_{3}$ flux vacua supported by both KK5$^{(1,a)}$/KK5$^{(2,i)}$-monopoles in Table~\ref{tab:DW-branes}. After a proper compact quotient, the internal manifold is constructed from the solvmanifold with one-form basis elements (\ref{one-form_integrated_general}). The simultaneous presence of the two types of KK$5$-monopoles makes it impossible to T-dualise this scenario to a geometric\footnote{Applying three T-dualities along, say, $\,\eta^{a}\,$ produces a non-geometric type IIA scenario with a non-geometric $Q$-flux of the type introduced in \cite{Shelton:2005cf}. The same holds if applying three T-dualities along $\,\eta^{i}$.} type IIA background with NS$5^{(1,a)}$- and NS$5^{(2,i)}$-branes. Therefore, this case can be viewed as a genuine type IIB scenario that  generically succeeds in stabilising the eight dilatons of the G$_{2}$-compactification in an AdS$_{3}$ flux vacuum. For the sake of concreteness, we focused on the simple flux choice in (\ref{gauge_fluxes_example_KK1&KK2}) and determined the set of $H$-functions (\ref{H-functions_color_full_KK1&KK2})-(\ref{HKK_scalings_full_KK1&KK2}) for the D1-D5-KK5 intersection that recovers the AdS$_{3}$ flux vacuum in the near-horizon region. The $H_{5}^{(4,5,6,7)}$-functions in (\ref{H-functions_flavour_full_KK1&KK2}) are non-harmonic, and a non-zero flux-induced tadpole for the $\widetilde{\textrm{O}5}/\widetilde{\textrm{D}5}^{(4,5,6,7)}$ brane sources (see Table~\ref{tab:sources}) appears in the AdS$_{3}$ flux vacuum. This comes along with a spectrum of low-lying dual operators of the form
\begin{equation}
\Delta = 
4 \,\, (\times 4) 
\,\,\,,\,\,\, 
3 \,\, (\times 4) \ ,
\end{equation}
which does not allow for the extremal or over-extremal arrangements $\,\Delta_{i} + \Delta_{j} = \Delta_{k} - 2 \, n\,$ with $\,n \in \mathbb{N}\,$, hence passing again the consistency constraint of \cite{Bobev:2025yxp}. 
\\

Taken together, the results presented in this work provide evidence that type IIB scale-separated AdS$_3$ flux vacua admit a higher-dimensional interpretation in terms of intersecting O-planes, D-branes and KK5-monopoles, although smearing effects are required for consistency. Several directions for future research remain open. First, since the scale-separated AdS$_{3}$ flux vacua under study require smeared O5-planes to cancel tadpoles, it would be interesting to investigate whether fully localised solutions exist building upon the results in, \textit{e.g.}, \cite{Junghans:2020acz,Junghans:2023yue,Emelin:2022cac,Emelin:2024vug}. Another interesting direction is to gain a better understanding of the CFT$_2$ duals, if any, of scale-separated AdS$_3$ flux vacua in type IIB string theory. In this regard, it would be particularly valuable to determine whether the D$1$-D${5}^{(1,2,3)}$ brane sector decouples from the bulk using the interpolating solution in Section~\ref{sec:interpolating_solution}, especially in light of the obstruction pointed out in \cite{Bedroya:2025ltj} and the recent results of \cite{Apers:2026lgi} concerning such a decoupling in the context of scale-separated DGKT vacua. Finally, it would also be interesting to extend the analysis presented here to other classes of scale-separated AdS vacua in string theory \cite{Miao:2025rgf,Tringas:2025bwe}. We hope to return to these questions in future work.

\section*{Acknowledgements}

We are grateful to Fien Apers, Nathan Bagshaw, Giuseppe Dibitetto, Carlos Hoyos, Patrick Meessen, and especially Vincent Van Hemelryck for many insightful and stimulating discussions. We also thank Matteo Morittu for past collaborations on related work. \'AA, AG and GS are partially supported by the grants from the Spanish government MCIU-22-PID2021-123021NB-I00 and MCIU-25-PID2024-161500NB-I00. The work of \'AA is also supported by the Severo Ochoa fellowship NAC-AT-PUB-ASV-2025 BP24-134. The work of C.R.-D is financed in part by the Coordenação de Aperfeiçoamento de Pessoal de Nível Superior (CAPES) and the Fundação de Amparo à Pesquisa e Inovação do Espírito Santo (FAPES). C.R.-D also acknowledges the Centro Brasileiro de Pesquisas Físicas (CBPF) and the University of Oviedo for their hospitality at the first stages of this work, and for providing a stimulating research environment. The work of GS is supported by “Angelo Della Riccia” Foundation.

\newpage

\appendix

\section{Type IIB supergravity in the string frame}
\label{app:10D_EOMs}

In the string frame, the bosonic sector of type IIB supergravity can be written in its democratic formulation \cite{Bergshoeff:2001pv} as
\begin{equation} 
\label{S_IIB_String_Frame}
\begin{array}{rcl}
2\kappa^{2} \, S_{\textrm{IIB}} &=& \displaystyle\int \textrm{d}^{10}X \, \sqrt{-G} \, \left[ e^{-2\Phi} \left(R^{(10)} \, + \, 4  (\partial\Phi)^{2} \, - \, \frac{1}{2 \cdot 3!} |H_{(3)}|^2 \right ) \, - \, \frac{1}{4} \sum_{p=0}^4 \frac{|F_{(2p + 1)}|^2}{(2p + 1)!}\right] \\[4mm]
& + &  \displaystyle\sum_{k=4}^{7} \tilde{S}^{(k)}_{\textrm{loc}} \,\, + \,\, \displaystyle\sum_{k=1}^{7} S^{(k)}_{\textrm{loc}} \, + \,\,  S^{\textrm{O}1/\textrm{D}1}_{\textrm{loc}} \ .
\end{array}
\end{equation}
The second line of this expression contains the contributions of the localised sources, grouped as: $\widetilde{\textrm{O}5}/\widetilde{\textrm{D}5}^{(k)}$ $(k=4, \dots, 7)$ of Table~\ref{tab:sources} (for which the non-trivial tadpoles in (\ref{Tadpole_O5/D5_4567}) can be induced by the fluxes), the domain-wall $\textrm{O}5/\textrm{D}5^{(k)}$ branes $(k=1, \dots, 7)$ and O$1$/D$1$, both presented in Table~\ref{tab:DW-branes}. The action for each source is the sum of WZ and DBI contributions \cite{Myers:1999ps}:
\begin{equation}
S_{\textrm{loc}} = S_{\textrm{WZ}} + S_{\textrm{DBI}} \, ,
\end{equation}
which for the present work will be expanded at zeroth-order in $\alpha'$. By recalling that \cite{Gimon:1996rq, Giveon:1998sr, Bergman:2001rp}
\begin{equation} \label{Orientifold_Tension}
T_{\mathrm{O}p^\pm} = \mu_{\mathrm{O}p^\pm} = \pm 2^{p-4} \, T_{\mathrm{D}p} \, 
\end{equation}
and considering O$p^-$-planes, the WZ terms for the three classes of sources can be written as:
\begin{equation}
\begin{array}{lclcl}
    \tilde{S}^{(k)}_{\textrm{WZ}} & = & \displaystyle\int_{10\textrm{D}}\,C_{(6)}\wedge\tilde{J}^{\widetilde{\textrm{O}5}/\widetilde{\textrm{D}5}}_{(k)} & = & -2\kappa^2 \, \big(2N_{\widetilde{\textrm{O}5}}^{(k)} - N_{\widetilde{\textrm{D}5}}^{(k)} \big) \, T_{\textrm{D}5} \, \displaystyle\int_{\textrm{WV($\widetilde{\textrm{O}5}$/$\widetilde{\textrm{D}5}^{(k)}$)}}\,C_{(6)}   \, , \\[6mm]
   S^{(k)}_{\textrm{WZ}} & = & \displaystyle\int_{10\textrm{D}}\,C_{(6)}\wedge J^{\textrm{O}5/\textrm{D}5}_{(k)} & = & -2\kappa^2 \, \big(2N_{\textrm{O}5}^{(k)} - N_{\textrm{D}5}^{(k)} \big) \, T_{\textrm{D}5} \, \displaystyle\int_{\textrm{WV($\textrm{O}5$/$\textrm{D}5^{(k)}$)}}\,C_{(6)}   \, ,
   \\[6mm]
    S^{\textrm{O}1/\textrm{D}1}_{\textrm{WZ}} & = & \displaystyle\int_{10\textrm{D}}\,C_{(2)}\wedge J^{\textrm{O}1/\textrm{D}1} & = & -2\kappa^2 \, \big(\frac18 N_{\textrm{O}1} - N_{\textrm{D}1} \big) \, T_{\textrm{D}1} \, \displaystyle\int_{\textrm{WV($\textrm{O}1$/$\textrm{D}1$)}}\,C_{(2)}   \, ,
\end{array}
\end{equation}
with $\tilde{J}^{\widetilde{\textrm{O}5}/\widetilde{\textrm{D}5}}_{(k)}$, $J^{\textrm{O}5/\textrm{D}5}_{(k)}$ and $J^{\textrm{O}1/\textrm{D}1}$  being the smearing functions for the different classes of branes:
\begin{equation}
\label{currents_branes}
    \tilde{J}^{\widetilde{
    \textrm{O}5}/\widetilde{\textrm{D}5}}_{(k)} \,  =\, \tilde{j}_{5(k)}\,\varphi_{(4)} \quad , \quad J^{
    \textrm{O}5/\textrm{D}5}_{(k)} \,  =\, j_{5(k)}\, dR\,\wedge\,\varphi_{(3)} \, \quad , \quad J^{\textrm{O}1/\textrm{D}1} \,  = \, j_{1}\,dR\,\wedge\,\varphi_{(3)}\,\wedge\,\varphi_{(4)} \, .
\end{equation}
The DBI terms, instead, are given by: 
\begin{equation}
\label{S_DBI_sources}
\begin{array}{lccl}
\tilde{S}^{(k)}_{\textrm{DBI}} & = &  2\kappa^2 \, \left (2N_{\widetilde{\textrm{O}5}}^{(k)} - N_{\widetilde{\textrm{D}5}}^{(k)} \right )  \, T_{\textrm{D}5} &  \displaystyle\int_{\textrm{WV($\widetilde{\textrm{O}5}$/$\widetilde{\textrm{D}5}^{(k)}$)}} d^6\sigma \ e^{- \Phi} \sqrt{- G^{(6\textrm{D})}_{(k)}} \ , \\[12 pt]
S^{(k)}_{\textrm{DBI}} & = &  2\kappa^2 \, \left (2N_{\textrm{O}5}^{(k)} - N_{\textrm{D}5}^{(k)} \right )  \, T_{\textrm{D}5} &  \displaystyle\int_{\textrm{WV($\textrm{O}5$/$\textrm{D}5^{(k)}$)}} d^6\sigma \ e^{- \Phi} \sqrt{- G^{(6\textrm{D})}_{(k)}} \ ,
\\[12 pt]
S^{\textrm{O}1/\textrm{D}1}_{\textrm{DBI}} & = &  2\kappa^2 \, \left (\frac18N_{\textrm{O}1} - N_{\textrm{D}1} \right )  \, T_{\textrm{D}1} & \displaystyle\int_{\textrm{WV ($\textrm{O}1$/$\textrm{D}1$)}} d^2\sigma \ e^{- \Phi} \sqrt{- G^{(2\textrm{D})}} \ ,
\end{array}
\end{equation}
with the integration of the smearing function already performed. Moreover, $ G^{(6\textrm{D})}_{(k)}\,$ is the determinant of the 10D metric pull-backed to the world-volume of the $k$-th O$5$/D$5$-source of any of the two classes. Similarly, $\,G^{(2\textrm{D})}\,$ denotes the determinant of the pullback of the 10D metric onto the world-volume of the O$1$/D$1$.

The pseudo-action \eqref{S_IIB_String_Frame} must be supplemented with the Hodge duality relations
\begin{equation}
\label{Hodge_Duality_Relations}
F_{(5)} = \star F_{(5)} 
\hspace{5mm} , \hspace{5mm}
F_{(7)} = - \star F_{(3)}
\hspace{5mm} , \hspace{5mm}
F_{(9)} = \star F_{(1)}
\hspace{5mm} , \hspace{5mm}
H_{(7)} = e^{-2\Phi}\,\star H_{(3)} \ ,
\end{equation}
and the (modified) Bianchi identities
\begin{equation}
\label{IIB_BI_Fluxes}
\begin{array}{rcl}
d H_{(3)} & = & 0\, ,\\[6pt]

d F_{(1)} & = & 0\, , \\[6pt]

d F_{(9)} \,-\, H_{(3)} \wedge F_{(7)} & = & 0\, , \\[6pt]
d F_{(5)} \,-\, H_{(3)} \wedge F_{(3)} & = & 0 \, , \\[6pt]
d H_{(7)} 
\,+\, \dfrac{1}{2}\displaystyle\sum_{p} \star F_{(p)}\wedge F_{(p-2)} & = & 0\, ,
\end{array}
\end{equation}
and
\begin{equation}
    \begin{array}{ccl}
    d F_{(3)} \,-\, H_{(3)} \wedge F_{(1)} & = & \displaystyle\sum_{k=4}^{7}\,\tilde{J}^{\widetilde{\textrm{O}5}/\widetilde{\textrm{D}5}}_{(k)} \,+\, \displaystyle\sum_{k=1}^{7}\,J^{\textrm{O}5/\textrm{D}5}_{(k)}\, ,\\[16pt]
    
    d F_{(7)} \,-\, H_{(3)} \wedge F_{(5)} & = & J^{\textrm{O}1/\textrm{D}1}\, .
    \end{array}
\end{equation}
The equation of motion of the dilaton picks a contribution from (\ref{S_DBI_sources}) and reads
\begin{equation}
\label{IIB_EOM_Dilaton}
\begin{array}{rcl}
2e^{-\Phi}\sqrt{-G}\left(R^{(10)}+4\,\Box\Phi-4\,(\partial\Phi)^2-\frac{|H_{(3)}|^2}{2\cdot3!}\right) + & &  \\[4mm]
- \,\, j_{1}\,\sqrt{- G^{(2\textrm{D})}}  
\,\,-\,\, \displaystyle\sum_{k=1}^{7}\,
j_{5(k)}\,\sqrt{- G^{(6\textrm{D})}_{(k)}}
\,\, - \,\,
\displaystyle\sum_{k=4}^{7}
\,\tilde{j}_{5(k)}\,\sqrt{- G^{(6\textrm{D})}_{(k)}} &=& 0 \ .
\end{array}
\end{equation}
Finally, Einstein's equation is given by
\begin{equation}
\label{IIB_Einstein_Eq}
\begin{array}{rcl}
\sqrt{-G}\,\left[e^{-2\Phi}
\left(
R_{MN}
+ 2\,\nabla_M \nabla_N\Phi
-  \dfrac{1}{2 \cdot 2!} (H_{(3)}^2)_{MN}
\right) \right. &&  \\[6mm]
- \dfrac{1}{2} (F_{(1)}^2)_{MN} 
- \dfrac{1}{2 \cdot 2!} (F_{(3)}^2)_{MN}
\left.- \dfrac{1}{4 \cdot 4!} (F_{(5)}^2)_{MN} 
+ \dfrac{1}{4} G_{MN}
\left(|F_{(1)}|^2 + \frac{1}{3!} |F_{(3)}|^2\right)\right] && \\[6mm]
+ j_{1}\,e^{-\Phi}\sqrt{- G^{(2\textrm{D})}}\left(\dfrac{1}{2}\,G^{(2\textrm{D})}_{MN} 
-\dfrac{1}{4}G_{MN}\right)&& \\[6mm]
+ \displaystyle\sum_{j=1}^{7}\,j_{5(k)}\,e^{-\Phi}\sqrt{- G^{(6\textrm{D})}_{(k)}}\left(\dfrac{1}{2}\,G^{(6\textrm{D})}_{(k)\,MN} 
-\dfrac{1}{4}G_{MN}\right)&& \\[6mm]
+ \displaystyle\sum_{j=4}^{7}\,\tilde{j}_{5(k)}\,e^{-\Phi}\sqrt{- G^{(6\textrm{D})}_{(k)}}\left(\dfrac{1}{2}\,G^{(6\textrm{D})}_{(k)\,MN} 
-\dfrac{1}{4}G_{MN}\right)&=& 0 \ ,
\end{array}
\end{equation}
which is obtained once ({\ref{IIB_EOM_Dilaton}}) is used to remove the Ricci scalar $\,R^{(10)}$. The above set of equations simplifies dramatically when $\,H_{(3)}=0\,$ and $\,F_{(1)}=F_{(5)}=0$, as required by the class of seven-dimensional manifolds with co-closed G$_{2}$-structure studied in this work.

\section{Examples of supersymmetric AdS$_{3}$ flux vacua}
\label{app:AdS3_vacua}

We will present various examples of supersymmetric AdS$_{3}$ flux vacua of the three-dimensional supergravity specified by the $\,\mathcal{N}=1\,$ superpotential in (\ref{W_model}). The first two families only activate one of the two metric flux types in (\ref{Maurer-Cartan_eq}), namely, either $\,\omega_{a}\,$ \textit{or} $\,\omega_{i}$. As argued in the main text, these two families are T-dual to type IIA compactifications on a straight seven-torus with D-branes and NS5-branes. However, while the first family succeeds in stabilising the moduli, the second one comes along with unstabilised moduli. The third family includes both types $\,\omega_{a}\,$ \textit{and} $\,\omega_{i}\,$ of metric fluxes simultaneously and is a variant of the type IIB flux vacua put forward in \cite{Arboleya:2024vnp,VanHemelryck:2025qok}. These three families serve as concrete examples to illustrate the general results presented in the main text.\footnote{In this appendix we have reinstated the explicit dependence of the three-dimensional solutions on the three-dimensional gauge coupling $\,g$.}

\subsection{AdS$_{3}$ vacua with KK5$^{(1)}$ monopoles}
\label{sec:AdS3_KK_1}

With the flux choice in (\ref{metric_fluxes_example_KK1}) and (\ref{gauge_fluxes_example_KK1}), the superpotential (\ref{W_model}) possesses a family of extrema which succeeds in stabilising the eight moduli fields. These extrema are located at
\begin{equation}
\label{VEVs_8_moduli_wtil=0}
\begin{array}{ccc}
\tau^{4} = -\left( \dfrac{1}{\omega^{8} \, \tilde{f}^{7}}\right) \, \left( f_{7} \, f_{31} \, f_{32} \, f_{33} \, f_{37} \right)^{3}
\hspace{8mm} , \hspace{8mm}
\rho^{28} = - \dfrac{f_{7}^{7}}{f_{31} \, f_{32} \, f_{33} \, \tilde{f}^{3} \, f_{37}} \ ,  \\[8mm]
\ell_{1}^{2} = - \left(\dfrac{\tilde{f}}{f_{37}}\right) \, \ell_{2}^{2} = -  \left(\dfrac{f_{31}}{\omega}\right)  \, \dfrac{\rho^{\frac{3}{2}}}{\tau^{\frac{1}{2}}} \ ,   \\[8mm]
\ell_{3}^{2} = - \left(\dfrac{\tilde{f}}{f_{37}}\right) \, \ell_{4}^{2} = -  \left(\dfrac{f_{32}}{\omega}\right)  \, \dfrac{\rho^{\frac{3}{2}}}{\tau^{\frac{1}{2}}} \ ,   \\[8mm]
\ell_{5}^{2} = - \left(\dfrac{\tilde{f}}{f_{37}}\right) \, \ell_{6}^{2} =  - \left(\dfrac{f_{33}}{\omega}\right)  \, \dfrac{\rho^{\frac{3}{2}}}{\tau^{\frac{1}{2}}} \ ,
\end{array}
\end{equation}
provided $\,\tilde{f}>0\,$ and $\,f_{37} < 0\,$. The vacuum energy at these supersymmetric AdS$_{3}$ flux vacua takes the form
\begin{equation}
\label{vacuum_energy_wtil=0}
V_{0} = - \dfrac{g^{2}}{2} \left(\dfrac{\omega^{6} \, \tilde{f}^{6}}{f_{37}^2}\right)  \left( \dfrac{1}{f_{7} \, f_{31} \, f_{32} \, f_{33}} \right)^{2}  \ ,
\end{equation}
which we will use to define the AdS$_{3}$ radius as $\,L^2 \equiv -\frac{2}{V_{0}}$. The normalised scalar masses and multiplicities read
\begin{equation}
\begin{array}{rcr}
m^2 L^2 &=& 
24 \,\, (\times 1) 
\,\,\,,\,\,\, 
8 \,\, (\times 7) \ , \\[2mm]
\Delta &=& 
6 \,\, (\times 1) 
\,\,\,,\,\,\, 
4 \,\, (\times 7) \ .
\end{array}
\end{equation}

\subsubsection*{Scales and string coupling}

The moduli VEV's in (\ref{VEVs_8_moduli_wtil=0}) set the characteristic size of the internal space ($L_{7}$) and the string coupling constant to be
\begin{equation}
L_{7} \equiv (\textrm{vol}_{7})^{\frac{1}{7}} =  \left( - \dfrac{f_{7}^{7}}{f_{31} \, f_{32} \, f_{33} \, \tilde{f}^3 \, f_{37}} \right)^\frac{1}{28}
\hspace{5mm} \textrm{ and } \hspace{5mm}
g_{s}^{2} = - \dfrac{f_{7} \, \tilde{f} \, \omega^2}{f_{31} \, f_{32} \, f_{33} \, f_{37} }  \ ,
\end{equation}
whereas the characteristic length of the three-dimensional external spacetime ($L_{3}$) reads
\begin{equation}
L_{3} \equiv \tau^{-1} \, L =   \dfrac{2}{g \, \omega} \,  \left( - \dfrac{f_{7} \, f_{31} \, f_{32} \, f_{33}\, f_{37}}{\tilde{f}^{5}} \right)^\frac{1}{4} \ . 
\end{equation}

\subsubsection*{Flux backtracking coefficients}

When backtracking these AdS$_{3}$ flux vacua, the three-dimensional domain-wall in (\ref{DW_sol_general}) has coefficients of the form
\begin{equation}
\label{DW_coeff_KK1}
c_{\tau}^{4} = (4 \, g)^3 \, \omega \,  \tilde{f}^2 
\hspace{5mm} , \hspace{5mm} 
c_{\rho} = \textrm{free}
\hspace{5mm} , \hspace{5mm} 
c_{\ell_{i}}^{24} = \left( \sqrt{-\dfrac{f_{37}}{\tilde{f}}} \,\, c_{\ell_{a}} \right)^{24} =   \dfrac{f_{37}^{8}}{4 \, g \, \omega^{3} \, \tilde{f}^{6}}\, c_\rho^4 \ .
\end{equation}

\subsection{AdS$_{3}$ vacua with  KK5$^{(2)}$ monopoles}
\label{sec:AdS3_KK_2}

The flux choice in (\ref{metric_fluxes_example_KK2}) and (\ref{gauge_fluxes_example_KK2}) produces a family of supersymmetric extrema with \textit{not} all the moduli being stabilised. In particular, from (\ref{general_AdS3_1})-(\ref{general_AdS3_2}) one finds
\begin{equation}
\label{VEVs_8_moduli_w=0}
\begin{array}{ccc}
\tau^{4} = \left( \dfrac{16 \, \kappa^{2}}{\omega^{8} \, f_{37}^{4}}\right)  \left[ f_{7} \, f_{31} \, f_{32} \, f_{33} \right]^{3}
\hspace{8mm} , \hspace{8mm}
\rho^{28} =  \dfrac{16 \, f_{7}^{7}}{f_{31} \, f_{32} \, f_{33} \, f_{37}^{4} \, \kappa^{6}} \ ,  \\[8mm]
\left(\dfrac{\ell_{1}}{\kappa}\right)^{2} = \ell_{2}^{2} = -\left(\dfrac{f_{31}}{\omega}\right) \, \dfrac{\rho^{\frac{3}{2}}}{\tau^{\frac{1}{2}}} \ ,   \\[8mm]
\left(\dfrac{\ell_{3}}{\kappa}\right)^{2} = \ell_{4}^{2} = -\left(\dfrac{f_{32}}{\omega}\right) \, \dfrac{\rho^{\frac{3}{2}}}{\tau^{\frac{1}{2}}} \ ,   \\[8mm]
\left(\dfrac{\ell_{5}}{\kappa}\right)^{2} = \ell_{6}^{2} = -\left(\dfrac{f_{33}}{\omega}\right) \, \dfrac{\rho^{\frac{3}{2}}}{\tau^{\frac{1}{2}}} \ .
\end{array}
\end{equation}
provided $\,f_{37} > 0$, and with an \textit{arbitrary} parameter $\,\kappa >0\,$ reflecting the presence of three unstabilised moduli. The vacuum energy at these supersymmetric AdS$_{3}$ flux vacua turns out to be independent of $\,\kappa\,$ and takes the form
\begin{equation}
\label{vacuum_energy_wtil=0}
V_{0} = - \dfrac{g^{2}}{32} \left(\omega^{6} \, f_{37}^{4} \right) \left( \dfrac{1}{f_{7} \, f_{31} \, f_{32} \, f_{33}} \right)^{2}  \ ,
\end{equation}
which we will use to define the AdS$_{3}$ radius as $\,L^2 \equiv -\frac{2}{V_{0}}$. The normalised scalar masses and multiplicities read
\begin{equation}
\begin{array}{rcr}
m^2 L^2 & = & 
48 \,\, (\times 1) 
\,\,\,,\,\,\, 
8 \,\, (\times 4) 
\,\,\,,\,\,\, 
0 \,\, (\times 3) \ , \\[2mm]
\Delta & = & 
8 \,\, (\times 1) 
\,\,\,,\,\,\, 
4 \,\, (\times 4) 
\,\,\,,\,\,\, 
2 \,\, (\times 3) \ .
\end{array}
\end{equation}
showing the presence of three massless directions.

\subsubsection*{Scales and string coupling}

The moduli VEV's in (\ref{VEVs_8_moduli_w=0}) set the characteristic size of the internal space ($L_{7}$) and the string coupling constant to be
\begin{equation}
L_{7} \equiv (\textrm{vol}_{7})^{\frac{1}{7}} =  2^{\frac{1}{7}}\left( \dfrac{f_{7}^{7}}{f_{31} \, f_{32} \, f_{33} \, f_{37}^4 \, \kappa^{6}   } \right)^\frac{1}{28}
\hspace{5mm} \textrm{ and } \hspace{5mm}
g_{s}^{2} = \dfrac{f_{7} \, \omega^{2}}{f_{31} \, f_{32} \, f_{33} \, \kappa^2 }  \ ,
\end{equation}
so they depend on the arbitrary parameter $\,\kappa$. Similarly, the characteristic length of the three-dimensional external spacetime ($L_{3}$) reads
\begin{equation}
L_{3} \equiv \tau^{-1} \, L =  \dfrac{4}{g \, f_{37} \, \omega \sqrt{\kappa}} \left(f_{7} \, f_{31} \, f_{32} \, f_{33} \right)^\frac{1}{4} \ . 
\end{equation}

\subsubsection*{Flux backtracking coefficients}

For this family of AdS$_{3}$ flux vacua, the coefficients entering the three-dimensional domain-wall in (\ref{DW_sol_general}) are given by 
\begin{equation}
\label{DW_coeff_KK2}
c_{\tau}^{3} = \frac{(4 \, g)^{2}  f_{37}^{2}}{4} \, \frac{c_{\rho}}{c_{\ell_{i}}^{6}} 
\hspace{5mm} , \hspace{5mm} 
c_{\rho} = \textrm{free}
\hspace{5mm} , \hspace{5mm} 
c_{\ell_{i}} = \textrm{free}
\hspace{5mm} , \hspace{5mm} 
c_{\ell_{a}}^{6}  =  \dfrac{f_{37}^{2}}{16 \, g \, \omega^{3}}\, \dfrac{c_\rho^4}{c_{\ell_{i}}^{18}} \ .
\end{equation}
From the charges in (\ref{q5_charges_interpolating}), the correct AdS$_{3}$ radius is recovered in (\ref{consistency_AdS_radius}) if the DW$_{3}$ free parameter $\,c_{\ell_{i}}\,$ is fixed such that
\begin{equation}
c_{\tau}^{4} = \dfrac{f_{37}^2 \, (4 \, g)^3 \, \omega}{4} \, \kappa^2
\hspace{8mm} , \hspace{8mm} 
\left(\frac{c_{\ell_{a}}}{\kappa} \right)^{24}  = c_{\ell_{i}}^{24} = \dfrac{f_{37}^{2}}{4 \, (4 \, g) \, \omega^{3}} \, \dfrac{c_\rho^4}{\kappa^{6}}  \ ,
\end{equation}
where $\,\kappa\,$ is the free parameter appearing in (\ref{VEVs_8_moduli_w=0}) due to the fact that this family of AdS$_{3}$ solutions comes along with unstabilised moduli.

\subsection{AdS$_{3}$ vacua with KK5$^{(1)}$ and KK5$^{(2)}$ monopoles}
\label{sec:AdS3_KK_1&KK_2}

The flux choice in (\ref{metric_fluxes_example_KK1&KK2}) and (\ref{gauge_fluxes_example_KK1&KK2}) produces a family of supersymmetric extrema which again succeeds in stabilising the eight moduli fields. These are located at
\begin{equation}
\label{VEVs_8_moduli_general}
\begin{array}{ccc}
\tau^{4} =  \dfrac{2^{-12}}{\tilde{f}^{4} \, \omega^{8}} \left[ f_{7} \, f_{31} \, f_{32} \, f_{33} \right]^{3}
\hspace{5mm} , \hspace{5mm}
\rho^{28} = \left(\dfrac{1}{16}\right)  \dfrac{f_{7}^{7}}{f_{31} \, f_{32} \, f_{33} \, \tilde{f}^{4}}  \ ,  \\[4mm]
\ell_{1}^{2} = \ell_{2}^{2} =  - \left( \dfrac{f_{31}}{2 \, \omega}\right) \, \dfrac{\rho^{\frac{3}{2}}}{\tau^{\frac{1}{2}}} \ , \\[4mm]
\ell_{3}^{2} =  \ell_{4}^{2} =  - \left( \dfrac{f_{32}}{2 \, \omega}\right) \, \dfrac{\rho^{\frac{3}{2}}}{\tau^{\frac{1}{2}}} \ , \\[4mm]
\ell_{5}^{2} = \ell_{6}^{2} =  - \left( \dfrac{f_{33}}{2 \, \omega}\right) \, \dfrac{\rho^{\frac{3}{2}}}{\tau^{\frac{1}{2}}} \ ,
\end{array}
\end{equation}
provided $\,\tilde{f} > 0$. The vacuum energy at these supersymmetric AdS$_{3}$ flux vacua takes the form
\begin{equation}
\label{vacuum_energy_general}
V_{0} = - g^{2} \, \left( 2^{9} \, \omega^{6} \, \tilde{f}^{4}\right) \, \left( \dfrac{1}{f_{7} \, f_{31} \, f_{32} \, f_{33}} \right)^{2} \ ,
\end{equation}
which we will use to define the AdS$_{3}$ radius as $\,L^2 \equiv -\frac{-2}{V_{0}}$. The normalised scalar masses and multiplicities read
\begin{equation}
\begin{array}{rcr}
m^2 L^2 & = & 
8 \,\, (\times 4) 
\,\,\,,\,\,\, 
3 \,\, (\times 4) \ , \\[2mm]
\Delta & = & 
4 \,\, (\times 4) 
\,\,\,,\,\,\, 
3 \,\, (\times 4) \ .
\end{array}
\end{equation}

\subsubsection*{Scales and string coupling}

The moduli VEV's in (\ref{VEVs_8_moduli_general}) set the characteristic size of the internal space ($L_{7}$) and the string coupling constant to be
\begin{equation}
L_{7} \equiv 2^{-\frac{1}{7}} \left( \dfrac{f_{7}^{7}}{f_{31} \, f_{32} \, f_{33} \, \tilde{f}^4} \right)^\frac{1}{28}
\hspace{3mm} \textrm{ and } \hspace{3mm}
g_{s}^{2} =\dfrac{4 \, \omega^{2} \, f_{7}}{f_{31} \, f_{32} \, f_{33}}  \ ,
\end{equation}
whereas the characteristic length of the three-dimensional external spacetime ($L_{3}$) reads
\begin{equation}
L_{3} \equiv \tau^{-1} \, L =  \dfrac{1}{ 2 \, g \, \omega \, \tilde{f}} \,  \left(f_{7} \, f_{31} \, f_{32} \, f_{33} \right)^\frac{1}{4} \ . 
\end{equation}

\subsubsection*{Flux backtracking coefficients}

For these supersymmetric AdS$_{3}$ flux vacua, the coefficients entering the three-dimensional domain-wall in (\ref{DW_sol_general}) take the form
\begin{equation}
\label{DW_coeff_general}
c_{\tau}^{4} = (8 \, g)^3 \, \omega \, \tilde{f}^2 
\hspace{6mm} , \hspace{6mm} 
c_{\rho} = \textrm{free}
\hspace{6mm} , \hspace{6mm} 
c_{\ell_{i}}^{24} =  c_{\ell_{a}}^{24} =  \dfrac{\tilde{f}^2}{ 8 \, g \, \omega^{3}}\, c_\rho^4 \ .
\end{equation}

\bibliography{references}

\end{document}